\documentclass[acmlarge]{acmart}

\usepackage[utf8]{inputenc}
\usepackage[T1]{fontenc}
\usepackage{hyperref}
\usepackage{epstopdf}

\usepackage{xcolor}
\usepackage{soul}

\usepackage{multirow}
\usepackage{tabularx} 
\usepackage[symbol]{footmisc}

\usepackage{mathtools, cuted}
\usepackage{algorithm}
\usepackage{algpseudocode}
\usepackage{caption}
\usepackage{balance}

\usepackage{subcaption} 
\usepackage{graphicx}

\usepackage{float}
\usepackage[switch]{lineno}
\usepackage{booktabs}  

\renewcommand{\textcolor}[2]{#2}

\AtBeginDocument{%
  }

\setcopyright{cc}
\setcctype{by-nc-sa}
\acmJournal{IMWUT}
\acmYear{2025} \acmVolume{9} \acmNumber{4} \acmArticle{205} \acmMonth{12} \acmPrice{}\acmDOI{10.1145/3770712}

\begin{document}

\title{Forecasting Occupational Survivability of Rickshaw Pullers in a Changing Climate with Wearable Data}


\author{Masfiqur Rahaman}\authornote{Corresponding author.}
\email{masfiq15@gmail.com}
\affiliation{%
  \institution{University of California San Diego}
  \city{San Diego}
  \state{California}
  \country{USA}
  \postcode{92093}
}

\author{Maoyejatun Hasana}
\email{hasana004@gmail.com}
\affiliation{%
  \institution{Bangladesh University of Engineering and Technology}
  \city{Dhaka}
  \country{Bangladesh}}

\author{Shahad Shahriar Rahman}
\email{2005092@cse.buet.ac.bd}
\affiliation{%
  \institution{Bangladesh University of Engineering and Technology}
  \city{Dhaka}
  \country{Bangladesh}}

\author{MD Sajid Mostafiz Noor}
\email{2005051@ugrad.cse.buet.ac.bd}
\affiliation{%
  \institution{Bangladesh University of Engineering and Technology}
  \city{Dhaka}
  \country{Bangladesh}}
  
\author{Razin Reaz Abedin}
\email{razin.reaz@gmail.com}
\affiliation{%
  \institution{Bangladesh University of Engineering and Technology}
  \city{Dhaka}
  \country{Bangladesh}}

\author{Md Toki Tahmid}
\email{1805030@ugrad.cse.buet.ac.bd}
\affiliation{%
  \institution{Bangladesh University of Engineering and Technology}
  \city{Dhaka}
  \country{Bangladesh}}

\author{Duncan Watson Parris}
\email{dwatsonparris@ucsd.edu}
\affiliation{%
  \institution{University of California San Diego}
  \city{San Diego}
  \state{California}
  \country{USA}
  \postcode{92093}
}

\author{Tanzeem Choudhury}
\email{tanzeem.choudhury@cornell.edu}
\affiliation{%
  \institution{Cornell University}
  \city{Ithaca}
  \state{New York}
  \country{USA}
}

\author{A. B. M. Alim Al Islam}
\email{alim_razi@cse.buet.ac.bd}
\affiliation{%
  \institution{Bangladesh University of Engineering and Technology}
  \city{Dhaka}
  \country{Bangladesh}}

\author{Tauhidur Rahman}
\email{trahman@ucsd.edu}
\affiliation{%
  \institution{University of California San Diego}
  \city{San Diego}
  \state{California}
  \country{USA}
  \postcode{92093}
}

\renewcommand{\shortauthors}{Rahaman et al.}

\begin{abstract}
While the vulnerability of cycle rickshaw pullers to extreme heat is well recognized, little effort has been devoted to modeling how their physiological biomarkers respond under such conditions. In this study, we collect real-time weather and physiological data using a wearable computing platform from 100 rickshaw pullers in Dhaka, Bangladesh. In parallel, we interview 12 additional rickshaw pullers to explore their knowledge, perceptions, and experiences related to climate change.
We propose a Linear Gaussian Bayesian Network (LGBN)-based regression model that predicts key physiological biomarkers based on activity, weather, and demographic features. The model achieves normalized mean absolute error (NMAE) of 0.82, 0.47, 0.65, and 0.67, respectively, for the biomarker: skin temperature, relative cardiac cost, skin conductance response, and skin conductance level.
Using climate model projections from 18 CMIP6 global climate models, we layer the LGBN on top of future climate forecasts to conduct a survivability analysis for both current (2023–2025) and future years (2026–2100). Based on the criteria $T_{WBGT} > 31.1^\circ$C and $T_{skin} > 35^\circ$C, the analysis shows that a significant percentage of rickshaw pullers (32\%) are already facing a high risk of heat-related illness or prolonged exposure to extreme heat ($T_{WBGT} > 31.1^\circ$C) during regular work hours. In future years, e.g., 2026-2030, based on the CMIP6-based climate models, this percentage can rise to 37 $\pm 17$\% with an exposure duration of 11.9 $\pm 2$ minutes (68\% of the trip duration) on average. A similar trend is found based on rickshaw pullers' skin temperature with exposure ($T_{skin} > 35^\circ$C) durations expanding from 11 minutes (64\% of the trip duration) to 13 ± 2 minutes (73\% of the trip duration) by 2026-2030.
Finally, a Thematic Analysis of interview data provides qualitative insights that complement the current observation and model’s predictions in the future. The findings reveal that rickshaw pullers acknowledge their growing climate vulnerability and express concern about its effects on their health and occupational survivability.
\end{abstract}

\ccsdesc[500]{Human-centered computing~Empirical studies in ubiquitous and mobile computing}
\ccsdesc[500]{Human-centered computing~Ubiquitous and mobile computing systems and tools}
\ccsdesc[500]{Human-centered computing~Ubiquitous and mobile computing design and evaluation methods}
\keywords{Climate Vulnerability, Climate Model, Occupational Survivability, Rickshaw Puller, Mobile and Wearable Sensing, Machine Learning and Artificial Intelligence}

\received{01 November 2024}
\received[revised]{05 August 2025}
\received[accepted]{18 September 2025}

\maketitle

\begin{figure}[!tbp]
    \centering    
    
\subfloat[A rickshaw \cite{rickshaw}]
{\includegraphics[width=.28\textwidth]{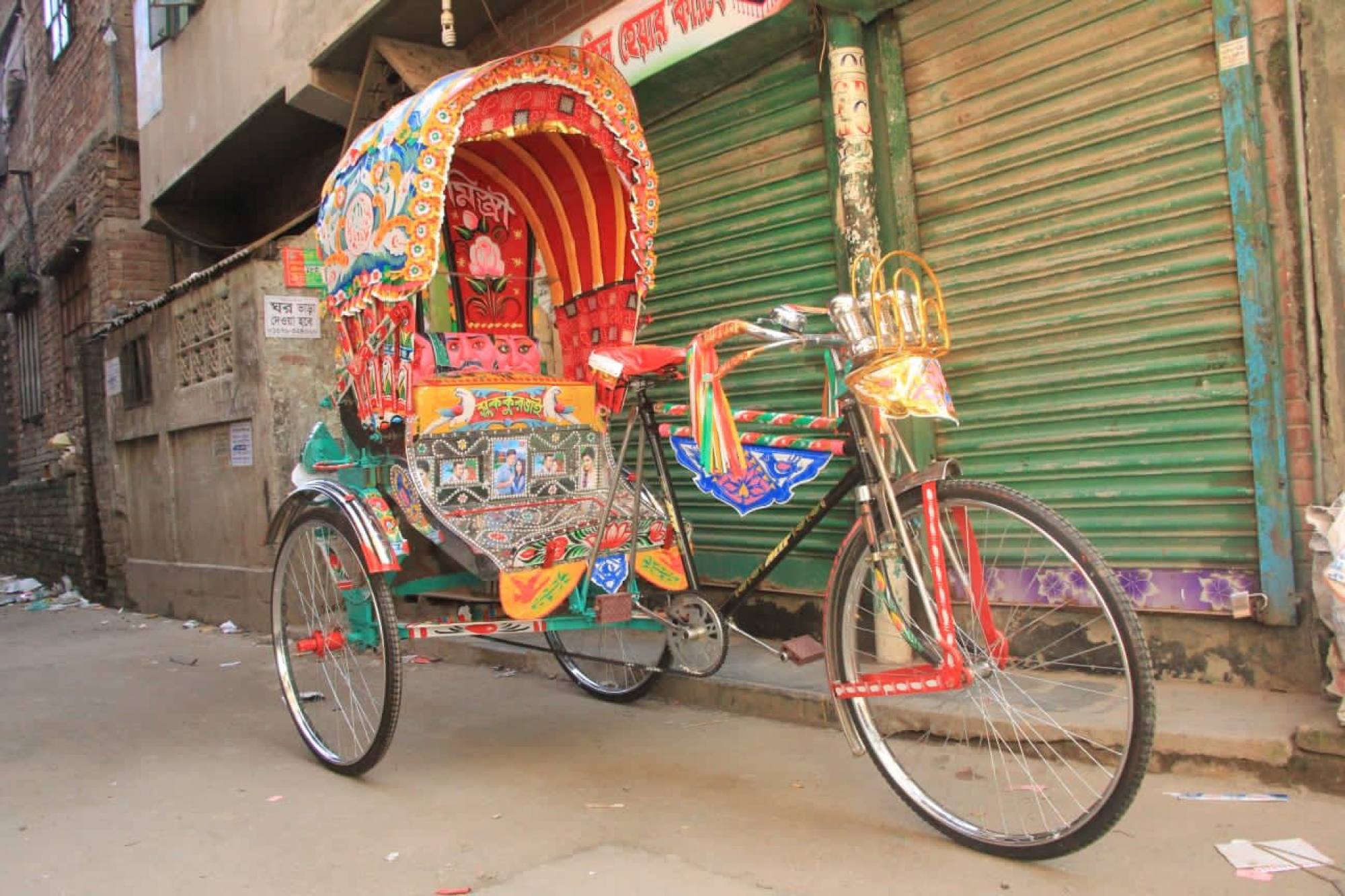}
\label{fig:rickshaw}}
\hfill
\subfloat[Data collection protocol]
{\includegraphics[width=.38\textwidth]{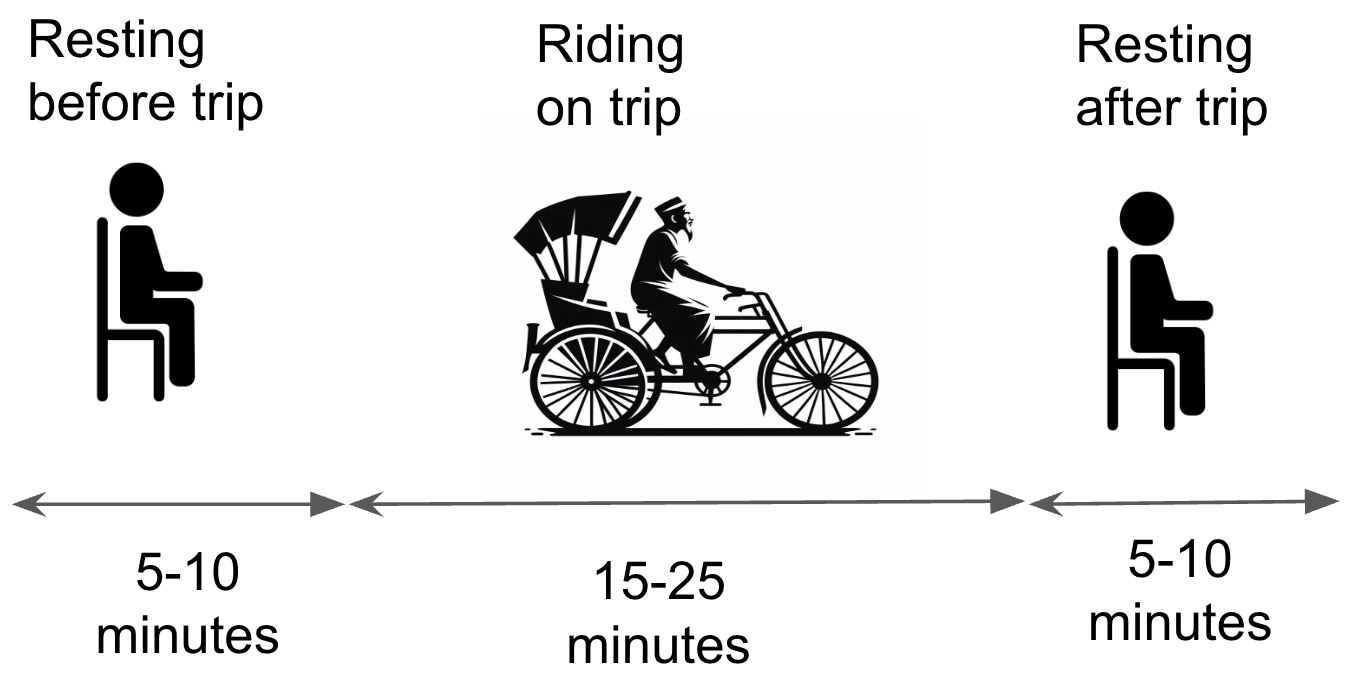}
\label{fig:data_col_protocol}}
\hfill
\subfloat[A rickshaw puller ]
{\includegraphics[width=.3\textwidth]{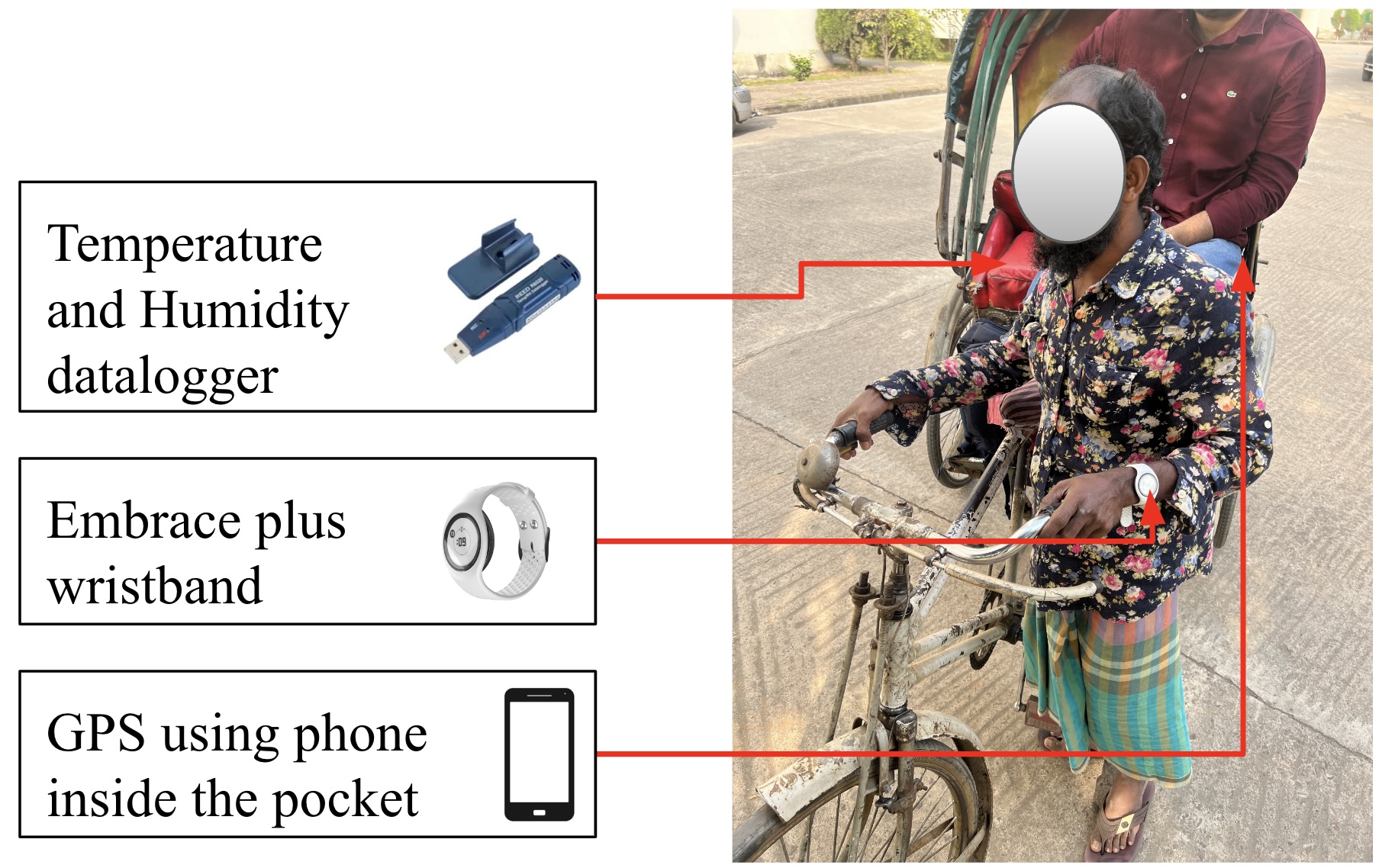}
\label{fig:data_collection}}

\caption{Data collection protocol with rickshaw pullers in Dhaka, showing the rickshaw, experimental timeline, and sensing devices (Embrace Plus wristband, temperature–humidity logger, and GPS smartphone).}
\end{figure}

\section{Introduction}
Extreme heat impacts people's livelihoods worldwide, particularly those in the urban global south. Daily laborers, who must perform manual labor in their workplaces, do not have the option to stop working on hot summer days. However, the increasing frequency of heat waves in many low and middle-income countries poses significant health risks, such as heat stroke and other heat stress-related illnesses \cite{heat_stroke}. Besides, wet bulb temperature is increasing and can exceed its critical threshold (35°C) for Southwest Asia by the end of this century \cite{Pal2016}.
The situation raises serious concerns about the survivability and well-being of daily laborers in this region. 

Rickshaw pullers, a specific group of daily laborers, predominantly found in urban areas of India, Bangladesh, etc., lead their livelihood by manually pulling a tricycle (i.e., rickshaw) on the streets with passengers and goods. 
The person who drives this tricycle is called a rickshaw puller.
Since cycle rickshaw pullers spend a prolonged time working outdoors, a rise in temperature due to climate change can have a significant detrimental impact on their health and well-being. Therefore, this study aims to understand the impact of heat exposure on several physiological biomarkers of rickshaw pullers.  

The necessity of this study is underscored by Bangladesh's vulnerability to climate change and the susceptibility of rickshaw pullers \cite{Kamal2024, heat_stroke}. Previous research has explored the heat impact of heat exposure on rickshaw pullers through questionnaires and limited-scale wearable sensing \cite{Chandan_2008, manna2012physiological, Sahu2013_tq}. \textcolor{red}{However, wearable sensing combined with the measurement of relevant biomarkers and weather conditions is essential to understand the physiological changes induced by heat stress.}

\begin{figure*}[!tbp]
    \centering
    \includegraphics[width=\linewidth]{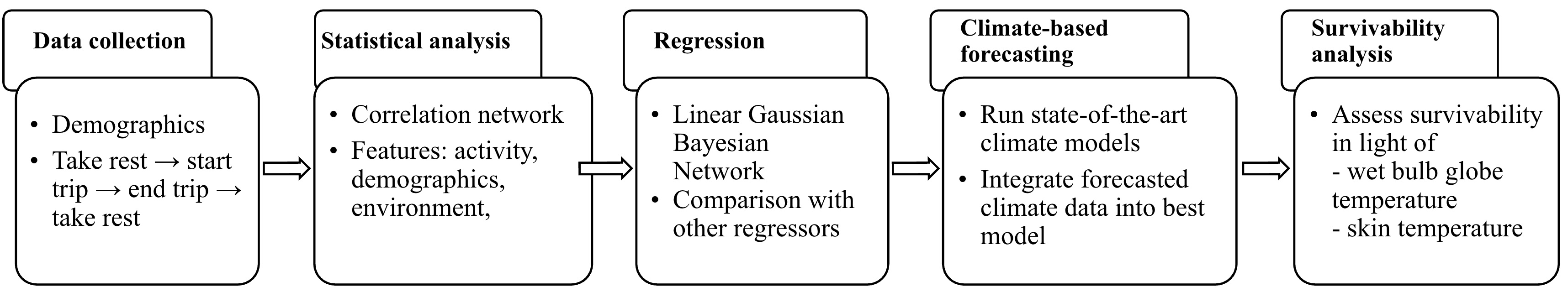}
    \caption{Our proposed methodology outlining each stage and subsequent tasks accomplished.}
\label{fig:methodology_flow_chart}
\end{figure*}

In this study, we employ a research-grade wearable device (e.g., Empatica Embrace Plus) \cite{embrace_plus, holt2020} for conducting a data collection effort on rickshaw pullers during their regular work schedules. The wristband is equipped with multiple sensors, including Photoplethysmography (PPG), electrodermal activity, accelerometer, and skin temperature. 
Besides, we measure the surrounding air temperature and relative humidity using the REED temperature and humidity data logger \cite{reed_instruments}.
Through performing necessary preprocessing on the collected data and feature extraction, we prepare a heat exposure dataset (n=100) of rickshaw puller participants.  
Through performing statistical and correlation-based analysis on the dataset as well as leveraging climate model forecasts and regressor models, we perform experiments to prepare a model that can forecast how physiological biomarkers (skin temperature, relative cardiac cost, skin conductance level, and skin conductance response) may change in future climate scenarios. 

\textcolor{red}{Based on the experimental analysis, between 2023–2025, one-third (32\%) of the rickshaw pullers in our study encountered a high risk of heat exposure ($T_{WBGT}$ > 31.1 °C).
CMIP6-based climate ensembles for mid-range warming (SSP245) suggest this share can reach 37 ± 17\% by 2026–2030 and 53 ± 15\% by 2091–2100. A similar pattern appears in skin temperature responses: the fraction of drivers whose skin temperature exceeds 35 °C climbs from 20\% (in 2023-2025) to 28 ± 18\% by 2026-2030, with exposure durations expanding from 11 minutes (64\% of the trip duration) to 13 ± 2 minutes (73\% of the trip duration). Together, these projections indicate a sharp escalation in occupational heat burden and survivability risk for rickshaw pullers over the coming decades.}


Furthermore, we interviewed an additional 12 participants, asking about their perception and experience with climate change. \textcolor{red}{Afterward, Thematic Analysis is performed on the collected interview scripts. The resultant themes are then compared with the outcomes of the aforementioned model. }


In this research, we particularly seek to answer the following questions:
\begin{itemize}
\item RQ1: What specific physiological changes do rickshaw pullers experience in response to heat stress during their regular work hours, and to what extent do demographics, physical activity, and weather affect the changes in physiological biomarkers?
\item RQ2: Can wearable sensing be used to forecast key physiological biomarkers (e.g., skin temperature, relative cardiac cost, skin conductance level, and skin conductance response) of the rickshaw pullers in understanding heat stress leveraging physiological and climate modeling?
\item RQ3: How survivable will the climate condition be for the rickshaw pullers in the coming years based on appropriate weather and physiological variables, such as wet bulb globe temperature, and skin temperature?
\end{itemize}

The remainder of the paper is structured as illustrated in Figure \ref{fig:methodology_flow_chart}. First, we describe our data collection methods, participant recruitment and management, and data preprocessing techniques. Next, we present the exploratory statistical analysis of the prepared dataset. Then, we describe regression models, highlighting a Linear Gaussian Bayesian Network and its comparison with conventional regressors. Finally, we discuss the concept of climate modeling, its use in forecasting various physiological features, and understanding the survivability of rickshaw pullers.


\begin{table}[!tbp]
\centering
\caption{Demographic factors and statistics of 100 subjects. The statistics are presented either in the form of mean and standard deviation (SD) values or count (n) and percentage values (\%).
}
\label{tab:demographics}
\begin{tabular}{|l|l|}
\hline
Factors & Statistics \\ \hline
\#Participants & 100 male participants; Interview on additional 6 participants \\ \hline
Season & Summer (\#participants = 51) \\
               & Winter (\#participants = 21) \\
               & Monsoon (\#participants = 28) \\
\hline
Trip duration (minutes), mean & 18.9 (5)\\ \hline
Weather & Mostly sunny, a few are cloudy and rainy \\ \hline
Age, mean(SD) & 48 (13) \\ \hline
BMI, mean (SD) & 20.6 (2.8) \\ \hline
Education & No formal education or below 5th grade \\ \hline
Daily income, mean & 6 USD \\ \hline
Total working hours/day, mean (SD) & 10 (2) \\ \hline
Duration of sleep (hrs)/day, mean (SD) & 6.8 (1.1) \\ \hline
\#Subjects having smoking habits & 56 \\ \hline
Disease & \begin{tabular}[c]{@{}l@{}}Cardiovascular (12), pain (51); Pain in hand, leg, knee, \\and head is reported\end{tabular} \\ \hline
\end{tabular}
\end{table}

\section{Dataset Preparation}
We developed a comprehensive dataset combining physiological biomarkers, weather conditions, activity level, and participant demographics. In the following subsections, we describe how the data were collected and prepared for analysis.
\subsection{Data Collection}
We recruited 100 rickshaw pullers from Dhaka, Bangladesh, for participation in data collection and an additional 12 rickshaw pullers to take part in structured interviews. Recruitment followed a random sampling approach, where participants were approached through face-to-face conversations on the streets and selected only if they were actively driving rickshaws in the city. Eligible participants were required to be at least 18 years old. Since rickshaw pullers in Dhaka are overwhelmingly male, the participant pool consisted entirely of men.

Given that low-income populations often have limited exposure to wearable sensing technologies, we first explained the study’s purpose, procedures, risks, and potential benefits in Bangla, the local language. Participants frequently asked follow-up questions, which we addressed to ensure clarity and understanding. The data collectors who are also co-authors of this study regularly use rickshaws for transportation and are familiar with the local dialect, enabling them to communicate effectively and build trust with participants.

In Figure \ref{fig:data_col_protocol}, we illustrate the data collection protocol. After recruitment, participants were asked to rest for 5–10 minutes, during which we collected demographic information such as age, height, weight, and food habits. Each participant was then fitted with an Embrace Plus wristband, which continuously recorded skin temperature, blood volume pulse, and electrodermal activity. In parallel, a temperature and humidity data logger measured ambient temperature and relative humidity. To capture spatial context, we used a smartphone and an additional GPS-enabled smartphone throughout the trip.

Once equipped, each participant completed a round-trip rickshaw ride between two city locations, lasting approximately 15–25 minutes, the typical trip length for rickshaw pullers in Dhaka\footnote{According to Hasan et al. \cite{HASAN2018246}, the average rickshaw trip in Dhaka is 3 km. With an average speed of 10 km/h, this corresponds to an average trip duration of about 18 minutes.}. After completing the ride, participants rested for another 5–10 minutes before concluding the session. A snapshot of a rickshaw puller during data collection is shown in Figure \ref{fig:data_collection}. Participants received compensation in addition to the standard fare for their trips.

All experimental protocols were reviewed and approved by the authors’ Institutional Review Board. Procedures were carried out in accordance with relevant ethical guidelines and regulations. At the start of the study, participants were informed about the study details, including their right to withdraw at any time and the expected duration of their involvement. Informed consent was obtained prior to participation.

\subsection{Data Preprocessing}

As participants were in motion during the data collection, noise from motion artifacts is present in the collected data. We apply different preprocessing steps on different data streams to mitigate or minimize the impact of motion artifacts.
In order to minimize the impact of motion artifacts on the Blood Volume Pulse (BVP) data, we use a second-order band-pass filter to retain frequencies in the range of 0.5 to 3 Hz, allowing for the capture of heart rates as high as 180 beats per minute, as observed during cycling activities \cite{nakagata2019heart}. Next, we address motion artifacts on the Electrodermal Activity (EDA) data. We visualize the power spectral density of the raw EDA signal for participants, followed by signal filtering using an 8th-order Butterworth low-pass filter with a cutoff frequency of 0.8 Hz. The choice of this filter is motivated by its similarity to a Chebyshev Type I filter used in previous studies \cite{Subramanian2019}.

\begin{table*}[!tbp]
\centering
\caption{Features extracted from the collected data designating type, description, and unit of measurements.}
\label{tab: features}
\resizebox{\linewidth}{!}{
\begin{tabular}{|l|l|l|l|}
\hline
Type & Feature & Description & Unit \\ \hline
\multirow{3}{*}{Weather} 
 & $T_{air}$ & Air temperature & ℃ \\ \cline{2-4} 
 & $R_H$ & Relative humidity & $\%$ \\ \cline{2-4} 
 & $T_{WBGT}$ & Wet bulb globe temperature & ℃ \\ \hline
 
\multirow{2}{*}{Demographics} 
 & $Age$ & Age of rickshaw puller & years \\ \cline{2-4} 
 & $BMI$ & Body Mass Index  & $kg/m^2$ \\ \cline{2-4} 
 & $Sleep$      & \textcolor{red}{Daily sleep duration}                         & hours \\ \cline{2-4}
& $t_{work}$ & \textcolor{red}{Daily work hours (pulling rickshaw outdoors)}      & hours \\ \hline
 \multirow{4}{*}{\parbox{2cm}{ Physiological \\ biomarkers}}& $RCC$ & \begin{tabular}[c]{@{}l@{}}Relative Cardiac Cost (RCC) refers to the percentage of heartbeats used\\ during rickshaw driving relative to resting condition.\end{tabular} & $\%$ \\ \cline{2-4} 
 & $SCR_n$& \begin{tabular}[c]{@{}l@{}}Skin Conductance Response ($SCR_n$) refers to the number of phasic \\(fast-changing) increase in the skin’s electrical conductance.\end{tabular}& $Count$\\  \cline{2-4} 
 & $SCL$ & \begin{tabular}[c]{@{}l@{}}Skin Conductance Level ($SCL$) is the tonic component of the electrodermal \\ activity of the skin which reflects the electrical conductance of the skin\end{tabular} & $\mu S$ \\ \cline{2-4} 
 & $T_{skin}$ & Wrist skin temperature of the rickshaw puller& ℃ \\ \hline

\multirow{2}{*}{Activity} & $Speed$ & Speed of the rickshaw during the trip & $km/h$ \\ \cline{2-4} 
 & $Dst_c$ & Distance traveled so far from the beginning to the current timestamp of the trip & $km$ \\ \cline{2-4} 
 & $Acc_m$ & \begin{tabular}[c]{@{}l@{}}Magnitude of acceleration experienced by the rickshaw puller during the trip, \\ taking into account acceleration in all three spatial dimensions (x, y, and z axes).\end{tabular} & $m/s^2$ \\ \cline{2-4} 
 & $t_{drive}$ & Duration of driving & $minutes$ \\ \hline
\end{tabular}
}
\end{table*}

\subsection{Feature Extraction}

Table \ref{tab: features} lists all 15 features used in our analysis. To ensure consistency across multimodal data streams sampled at different frequencies (e.g., weather, geolocation, and physiological signals), we adopt a window-based extraction strategy. Specifically, we apply a one-minute sliding window to all data streams. This choice follows ISO 9886, which recommends averaging heart rate over one-minute intervals \cite{ISO9886_2004}. Within each window, we compute weather, activity, physiological, and demographic features, described below.

Weather variables include air temperature, relative humidity, and solar radiation. Air temperature and relative humidity are collected at 1 Hz and averaged within each one-minute analysis window. Solar radiation data come from the ERA5 reanalysis product, which provides hourly averages; these values are aligned with the corresponding time and day of field data collection \cite{hersbach2023era5}. Using these variables, we estimate the wet bulb globe temperature ($T_{WBGT}$) following Equations \ref{eq:globe} and \ref{eq:wet_bulb_globe} \cite{Kamal2024}.

Activity features derive from accelerometer and GPS data. Accelerometer signals are collected at 64 Hz, and we compute the magnitude of acceleration ($Acc_m$) from the x, y, and z axes. GPS data are recorded at 1 Hz and averaged within each one-minute window to obtain the duration of rickshaw pulling ($t_{drive}$), average speed ($Speed$), and cumulative distance traveled ($Dst_c$).

Physiological biomarkers derive from Blood Volume Pulse (BVP), Electrodermal Activity (EDA), and skin temperature signals. BVP is recorded at 64 Hz, and we extract heart rate using 10-second windows with a one-second stride. We then estimate relative cardiac cost ($RCC$) from working heartbeats, resting heartbeats, and duration of rickshaw pulling(Equations \ref{eq:ncc} and \ref{eq:rcc}) \cite{Chandan_2008, Sahu2013_tq, hasana2025}.  
EDA is collected at 4 Hz. From this signal, we extract both tonic and phasic components. The skin conductance level ($SCL$), representing the tonic component, comes from the NeuroKit library \cite{Makowski2021neurokit}, while skin conductance responses (SCRs) correspond to the number of rapid, short-lived increases in the phasic component.  
Skin temperature is recorded at 1 Hz, and we compute its one-minute average within each analysis window to capture thermal responses during rickshaw pulling.

Demographic features include static participant characteristics that complement the time-varying weather, activity, and physiological measures. These variables are age ($Age$, years), body mass index ($BMI$, $kg/m^2$), daily sleep duration ($Sleep$, hours), and daily work hours spent pulling rickshaw outdoors ($t_{work}$, hours).


\begin{equation}
%
%
T_{g} = 0.009624 \cdot SR + 1.102 \cdot T_{air} - 0.00404 \cdot R_H - 2.2776
\label{eq:globe}
\end{equation}

\begin{equation}
\begin{aligned}
T_{WBGT} = 0.7 \cdot T_W + 0.1 \cdot T_{air} + 0.2 \cdot T_g
\end{aligned}
\label{eq:wet_bulb_globe}
\end{equation}

\begin{equation}
\begin{aligned}
NCC &= \text{Sum of working heart beats} - \text{resting heart beats per minute}  \cdot (\text{period of cycling in minutes})
\end{aligned}
\label{eq:ncc}
\end{equation}

\begin{equation}
RCC = \left(\frac{NCC}{(HR_{\text{max}} - HR_{\text{rest}}) \cdot \text{working period}}\right) \cdot 100
\label{eq:rcc}
\end{equation}

\begin{figure*}[!tbp]
\centering

\subfloat[]
{\includegraphics[width=0.23\textwidth]{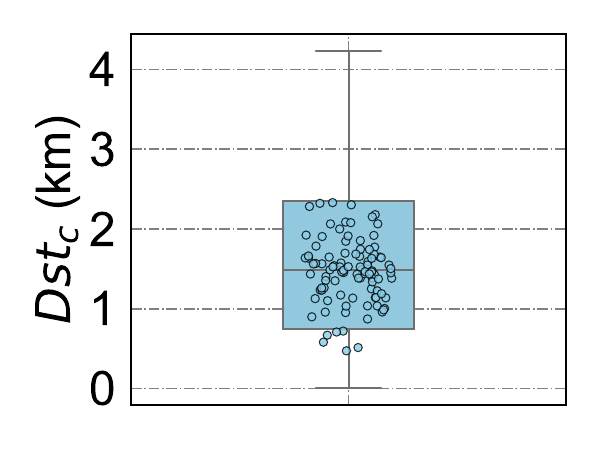} \label{fig:a_data_description_trip_Dst_c}}
\hspace{2 pt}
\subfloat[]
{\includegraphics[width=0.23\textwidth]{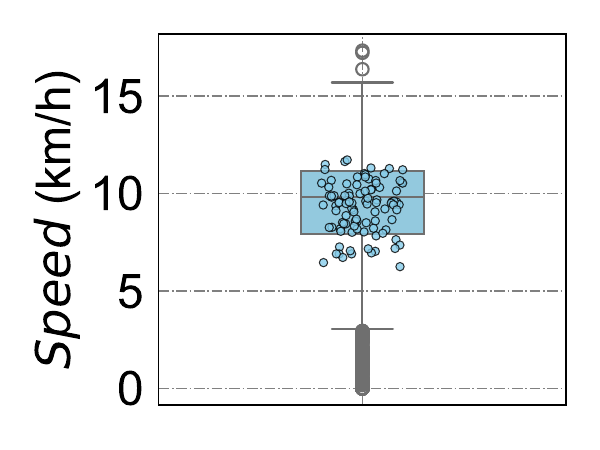} \label{fig:a_data_description_trip_Speed}}
\hspace{2 pt}
\subfloat[]
{\includegraphics[width=0.23\textwidth]{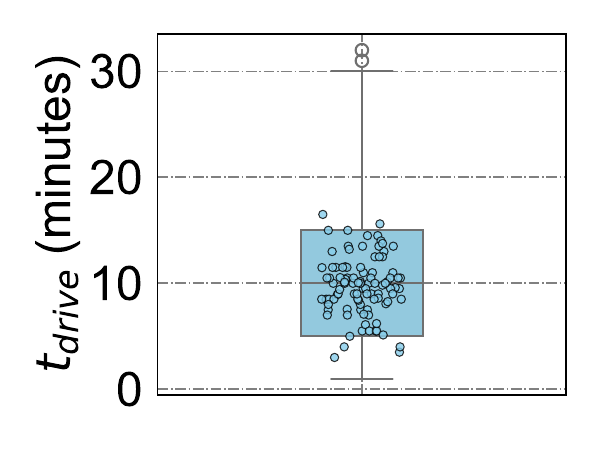} \label{fig:a_data_description_trip_t_drive}}

\subfloat[]
{\includegraphics[width=0.23\textwidth]{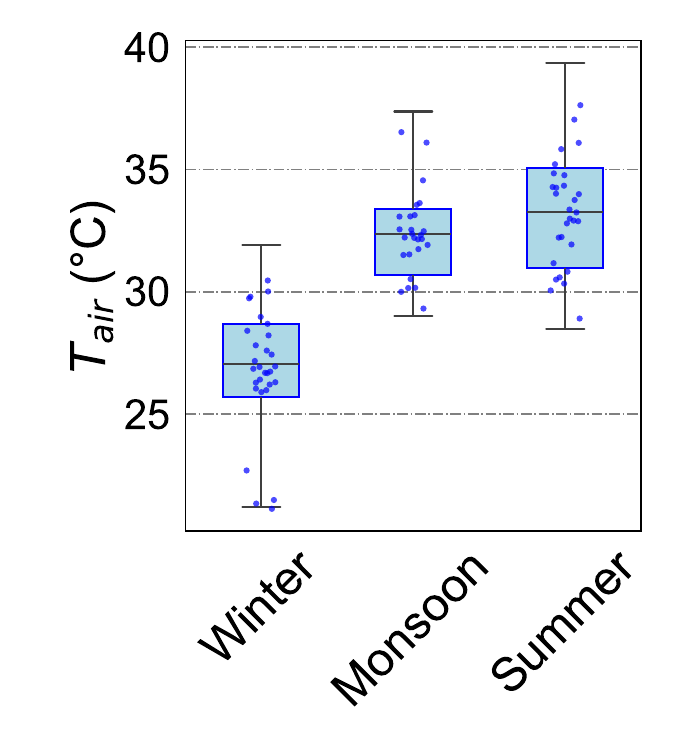} \label{fig:a_data_description_env_T_air}}
\hfill
\subfloat[]
{\includegraphics[width=0.23\textwidth]{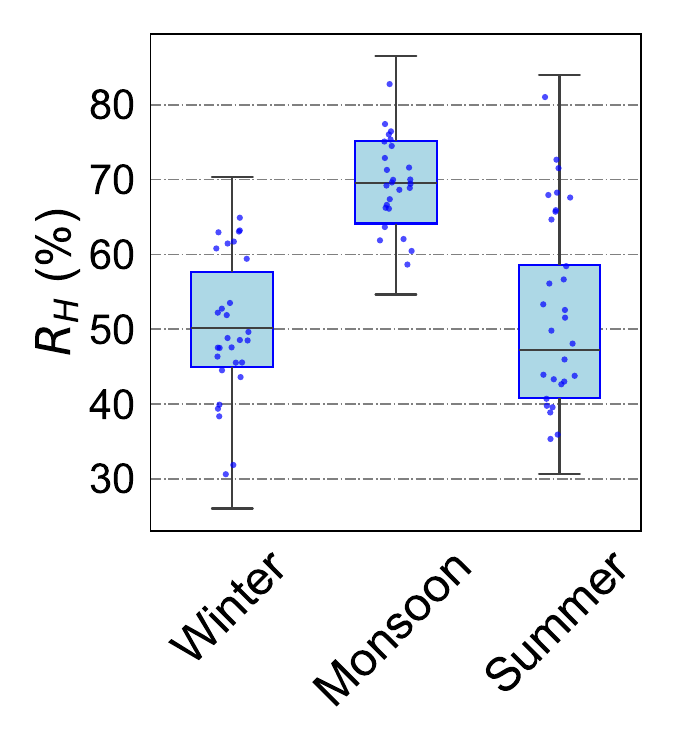} \label{fig:a_data_description_env_R_H}}
\hfill
\subfloat[]
{\includegraphics[width=0.23\textwidth]{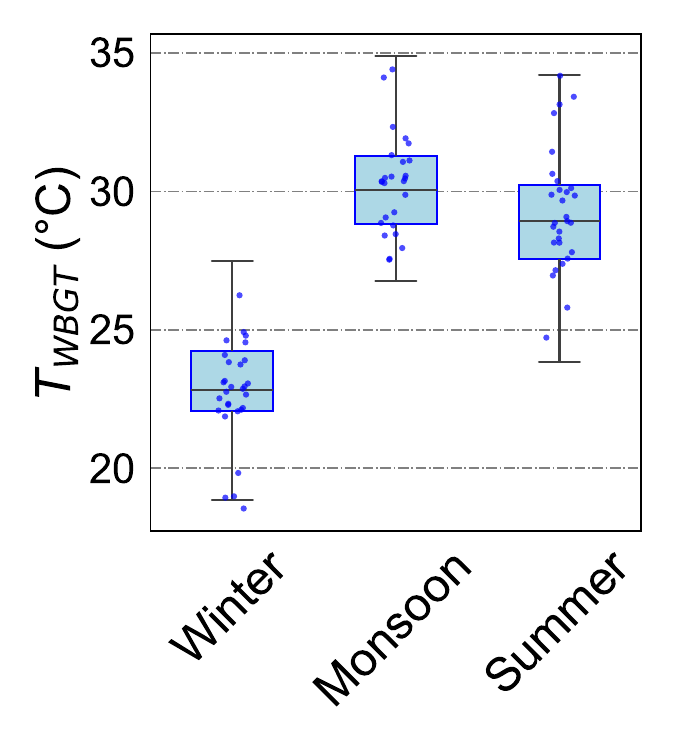} \label{fig:a_data_description_env_T_WBGT}}
\hfill
\subfloat[]
{\includegraphics[width=0.23\textwidth]{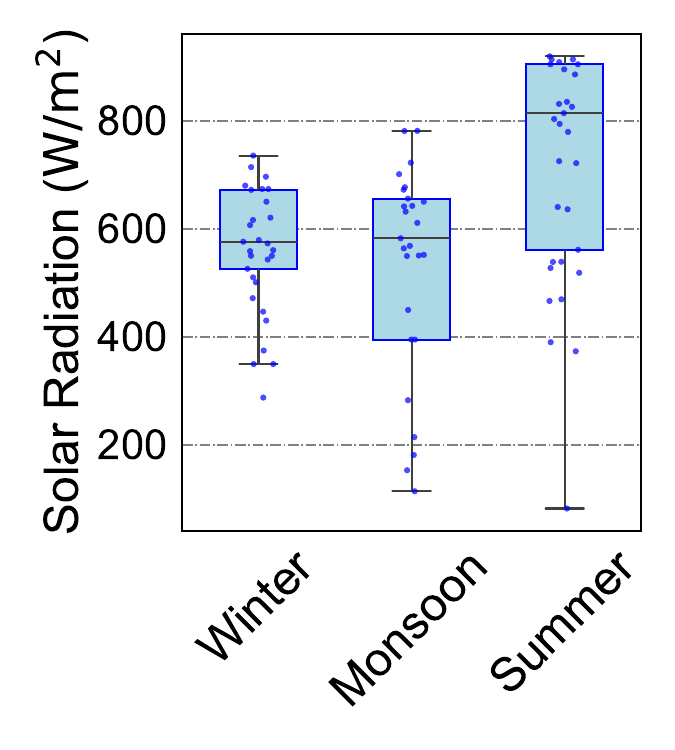} \label{fig:a_data_description_env_Solar_Radiation}}

\subfloat[]
{\includegraphics[width=0.23\textwidth]{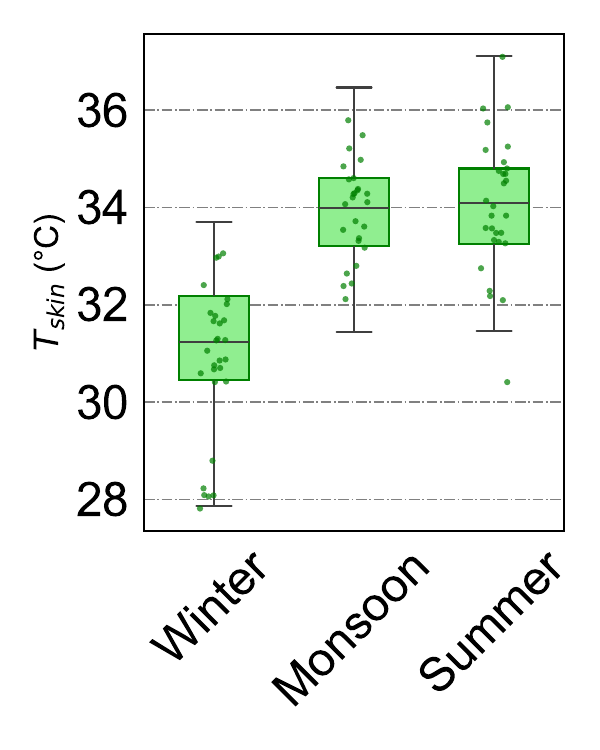} \label{fig:a_data_description_env_T_skin}}
\hfill
\subfloat[]
{\includegraphics[width=0.23\textwidth]{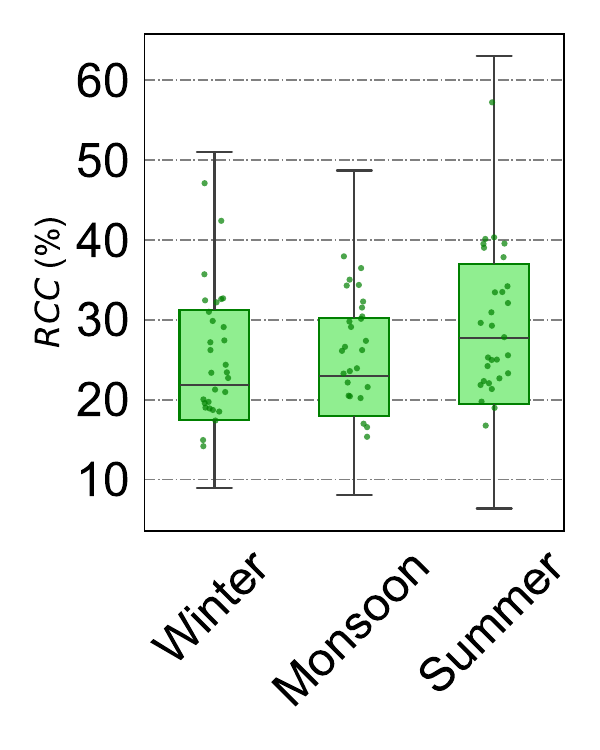} \label{fig:a_data_description_env_RCC}}
\hfill
\subfloat[]
{\includegraphics[width=0.23\textwidth]{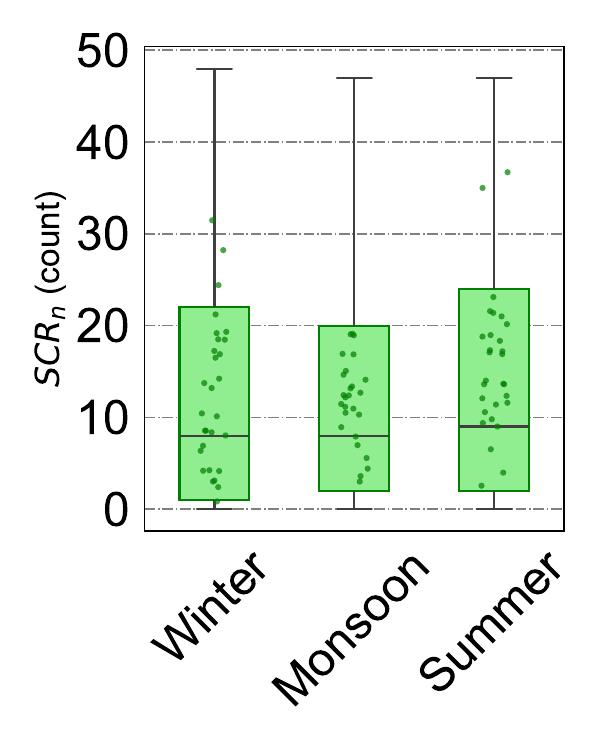} \label{fig:a_data_description_env_SCR_n}}
\hfill
\subfloat[]
{\includegraphics[width=0.23\textwidth]{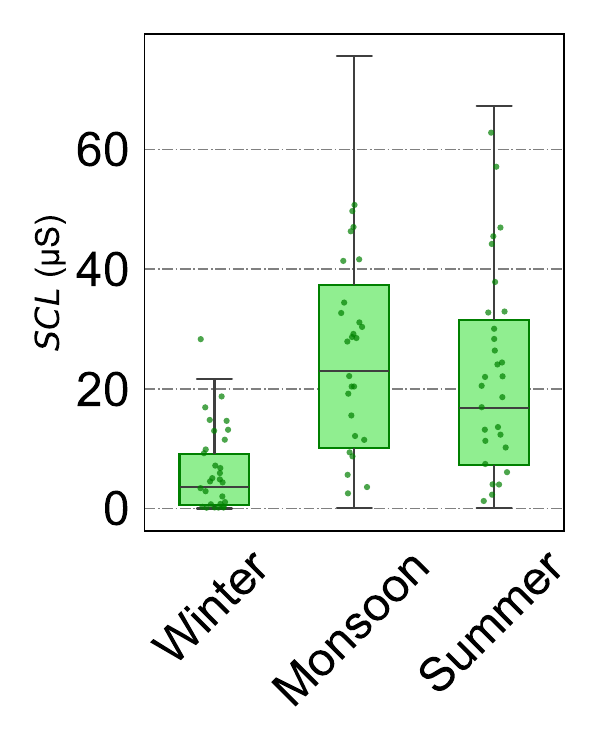} \label{fig:a_data_description_env_SCL}}

\subfloat[]
{\includegraphics[width=0.23\textwidth]{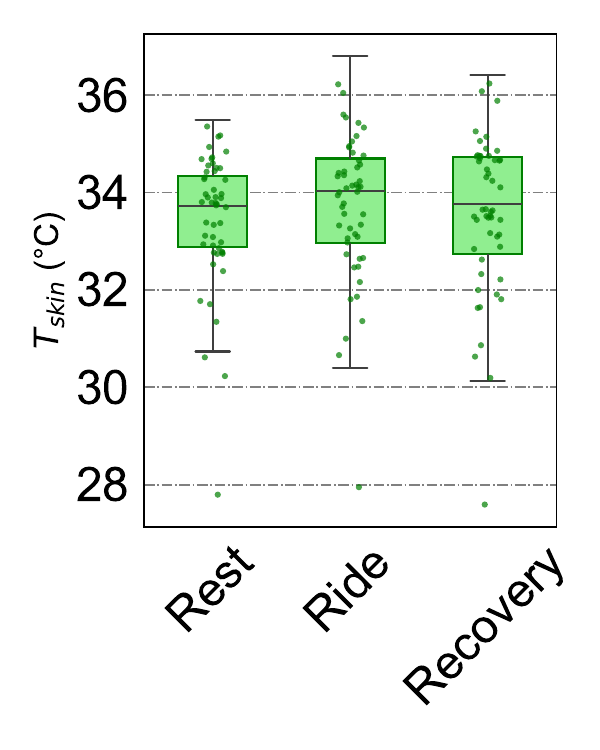} \label{fig:a_data_description_T_skin}}
\hfill
\subfloat[]
{\includegraphics[width=0.23\textwidth]{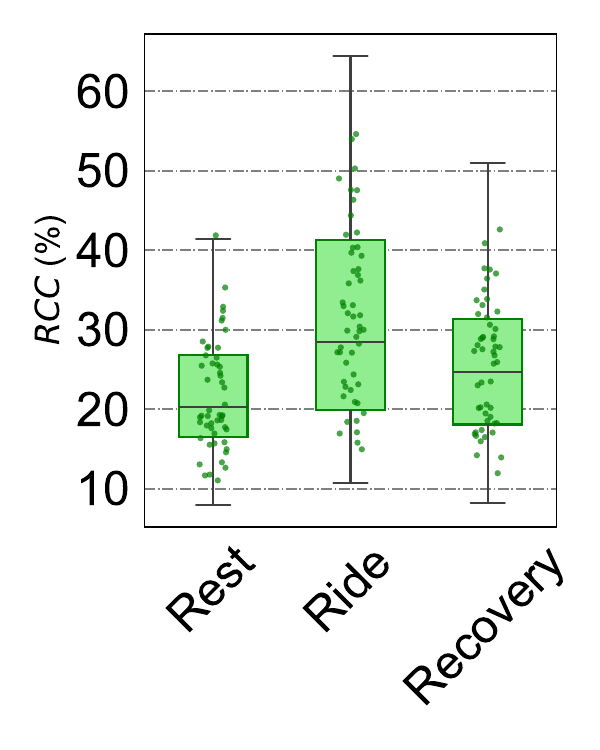} \label{fig:a_data_description_RCC}}
\subfloat[]
{\includegraphics[width=0.23\textwidth]{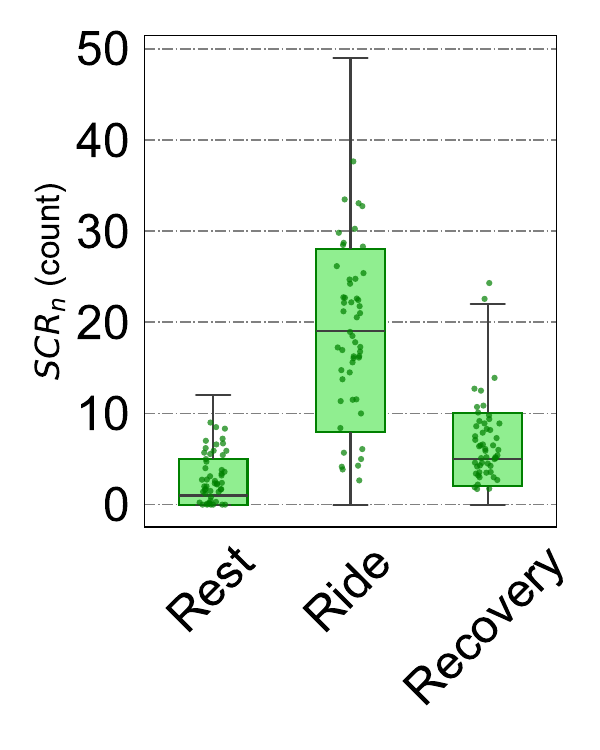} \label{fig:a_data_description_SCR_n}}
\hfill
\subfloat[]
{\includegraphics[width=0.23\textwidth]{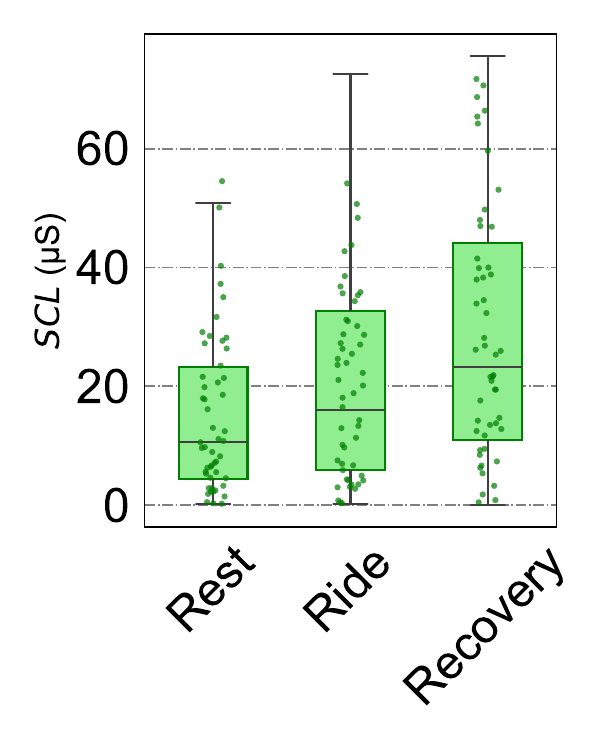} \label{fig:a_data_description_SCL}}
\hfill

\caption{Activity, Environmental and Physiological variables in three seasons and different stages of data collection. Small circles denote averages of the variables for each subject.}
\label{fig:data_description}
\end{figure*}

\subsection{Description of Data}
Our dataset comprises 100 rickshaw pullers' resting, riding, and recovery data for 53 hours and 18 minutes in total. 
The demographic characteristics of the participants are summarized in Table \ref{tab:demographics}. All participants are male with an average age of 48 (std=14) years. During our data collection, the duration of each rickshaw trip was 18.9 (std=5) minutes on average. The prevailing weather conditions during data collection are mostly sunny. Among the collected data, 51 rickshaw pullers participated during the Summer, 21 participated during the Winter, and 28 participated during the Monsoon of 2024 \footnote{Mid-June to mid-October are considered Monsoon in Bangladesh as heavy rainfall happens during this time. Though during Monsoon, there is heavy rain in Bangladesh, it is hot and humid as well which is also evident from Figure \ref{fig:a_data_description_env_R_H} and \ref{fig:a_data_description_env_T_air}}. A significant proportion of the participants have limited educational qualifications. Many participants report various types of pain associated with the physically demanding nature of their work. In Figure \ref{fig:data_description}, we present an overview of the activity level, environmental conditions, and physiological biomarkers during the data collection.

\subsection{Physiological Significance}

\textcolor{red}{To characterize overall physical strain, we focus on three physiological axes: thermoregulation, cardiovascular load, and sweat activation. As core temperature measurements are impractical during rickshaw operation, we rely on skin temperature ($T_{skin}$) as a non-invasive alternative. Cardiac load is assessed with Relative Cardiac Cost (RCC), a widely used index of occupational effort \cite{Chaudhuri2023, ijerph19137695}. Finally, electrodermal activity captures sweat gland output: the tonic skin-conductance level ($SCL$) and the phasic skin-conductance responses per minute ($SCR_n$), both established markers of sympathetic arousal \cite{s23083984, ISHIKAWA2025100877}.}

\textcolor{red}{Building on physiological foundations, we note that a healthy individual typically maintains a core body temperature of 37 ℃, which varies slightly based on individuals. On the other side, skin temperature is regulated at or below 35 ℃ in normal conditions to ensure heat dissipation \cite{Sherwood2010}. Sustained skin temperature above 37-38 ℃ can elevate core temperature to lethal values (42-43 ℃) even for fit individuals \cite{Sherwood2010}. Based on this, we categorize skin temperature into low (15–30℃), normal (30–35℃), and high (>35℃). The same categorization is also used in a clinical research study by Wilson et al., \cite{1_WILSON2025100851}.
Next, occupational physiology literature identifies 30\% relative cardiac cost ($RCC$) as a benchmark for moderate cardiovascular strain, with >40\% representing physiological overload and a potential trigger for acute cardiovascular events during extended workloads in hot environments \cite{6_Chaudhuri}. We use a similar categorization of RCC. It defines effort levels as follows: 0–19\% RCC indicates light effort, 20–39\% reflects moderate effort, and >40\% is considered vigorous or heavy effort \cite{5_s150716956}. Further, skin Conductance Response ($SCR_n$) reflects sympathetic nervous system activity. We follow Boucsein \cite{7_Boucsein2012} in categorizing <4 peaks/min as resting, 4–20 as moderate arousal, and >20 as high arousal \cite{7_Boucsein2012}.  Besides, Skin Conductance Level ($SCL$) values up to 20 $\mu S$ are considered resting to moderate stress, while values >20 $\mu S$ indicate extreme sympathetic arousal \cite{7_Boucsein2012, 9_Dawson2016, 10_Christie}. }

\begin{figure}[!tbp]
    \centering
    \includegraphics[width=0.9\linewidth]{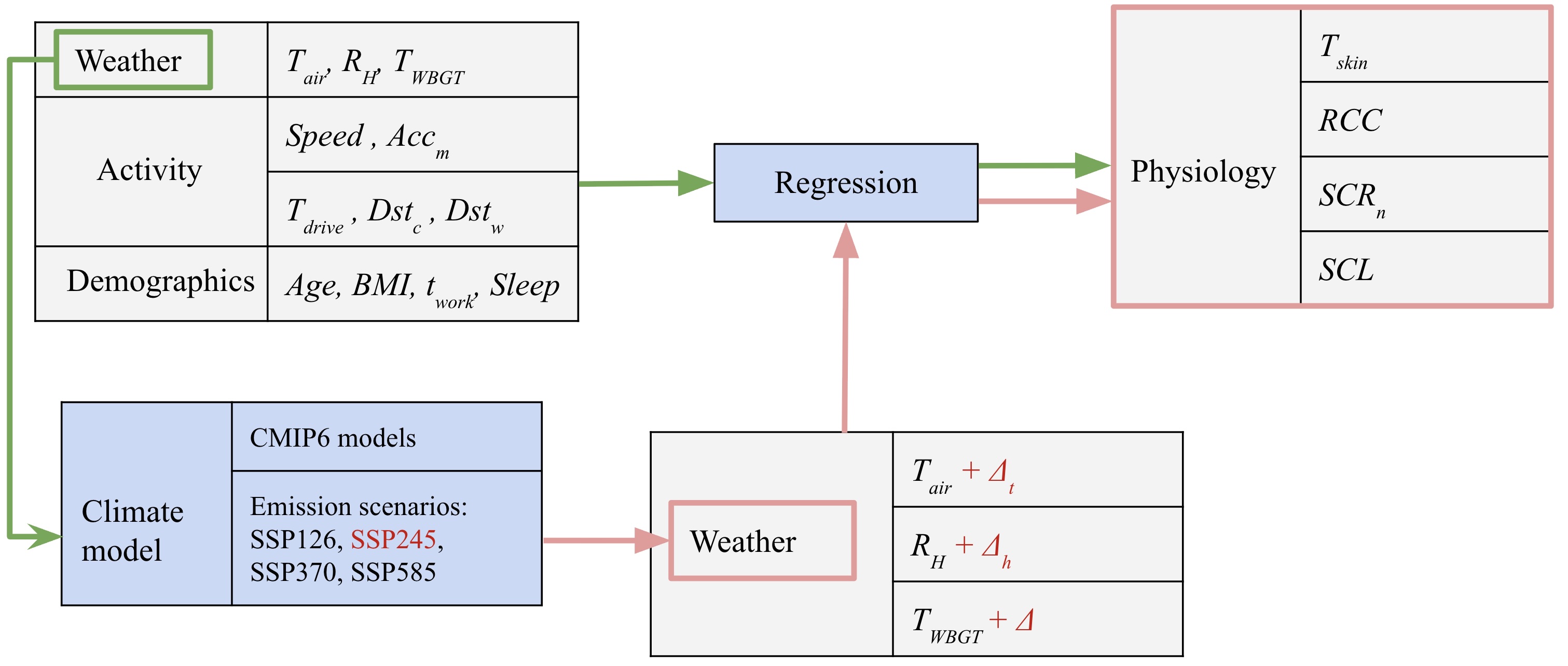}
    \caption{Modeling physiological biomarkers based on weather, season, and activity as well as integration of climate model output into the model.}
    \label{fig:ml_and_climate}
\end{figure}

\section{Correlation Network of Activity, Weather, and Physiological Biomarkers}
This section introduces a correlation network that captures how activity, weather, and demographics are associated with physiological biomarkers, providing a foundation for subsequent analysis.
\subsection{Method}
To explore the relationship among activity, weather, demographics, and physiological biomarkers, we construct a correlation network based on pairwise correlation. We utilize the Pearson correlation coefficient from the Scipy stats package \cite{pearsonr_scipy}, to compute coefficients and associated p-values for each pair of features.


In preparing the correlation network, we only consider the correlation between the following pairs: activity and physiological biomarkers; season and physiological biomarkers; weather and physiological biomarkers; and physiological biomarkers themselves. As the focus of this study is to understand the relationship between physiological biomarkers and other variables, we do not focus on the full cross-correlation matrix. 
Next, we adjust the p-values of the correlation matrix by the Benjamini-Hochberg procedure \cite{Benjamini_Hochberg} and consider statistically significant correlations (i.e., coefficient > 0.1 and p-value < 0.05) to remove spurious or false positive correlations \cite{AKOGLU201891, epskamp2018tutorial}. 

\begin{table}[]
\centering
\caption{Correlation matrix of the corresponding correlation network in Figure \ref{fig:correlation_netoworks}. Significantly strong correlations (Coefficient \textgreater= 0.1 and P $<$ 0.05) are shown with asterisks ($^*$)}
\label{tab:correlation_matrix}
\begin{tabular}{lcccccccccccccc}
\toprule
         & $T_{skin}$ & $RCC$ & $SCR_n$ & $SCL$ & $Speed$ & $Dst_c$ & $t_{drive}$ & $T_{WBGT}$ & $T_{air}$ & $R_H$ & $Age$ & $BMI$ & $t_{work}$ & $Sleep$\\
\midrule
$T_{skin}$  &             &       &         &         & 0.01 & 0.02 & 0.02 & 0.82$^*$ & 0.85$^*$ & 0.00 & -0.08 & -0.28 & 0.14$^*$ & -0.05\\
$RCC$       & 0.26$^*$        &       &         &         & 0.24$^*$ & 0.12$^*$ & 0.08 & 0.17$^*$ & 0.23$^*$ & -0.15 & 0.28$^*$  & 0.15$^*$  & 0.16$^*$ & -0.04\\
$SCR_n$     & 0.19$^*$        & 0.15$^*$  &         &         & 0.14$^*$ & 0.24$^*$ & 0.23$^*$ & 0.27$^*$ & 0.29$^*$ & -0.01 & -0.08 & -0.13 & 0.06 & -0.09\\
$SCL$       & 0.40$^*$        & 0.26$^*$  & 0.41$^*$    &         & 0.03 & 0.26$^*$ & 0.23$^*$ & 0.53$^*$ & 0.49$^*$ & 0.18$^*$  & -0.04 & 0.00  & 0.07 & -0.13$^*$\\
\bottomrule
\end{tabular}
\end{table}

\begin{figure*}[!tbp]
\centering
\subfloat[Activity and physiological \\biomarkers]
{\includegraphics[width=0.33\textwidth]{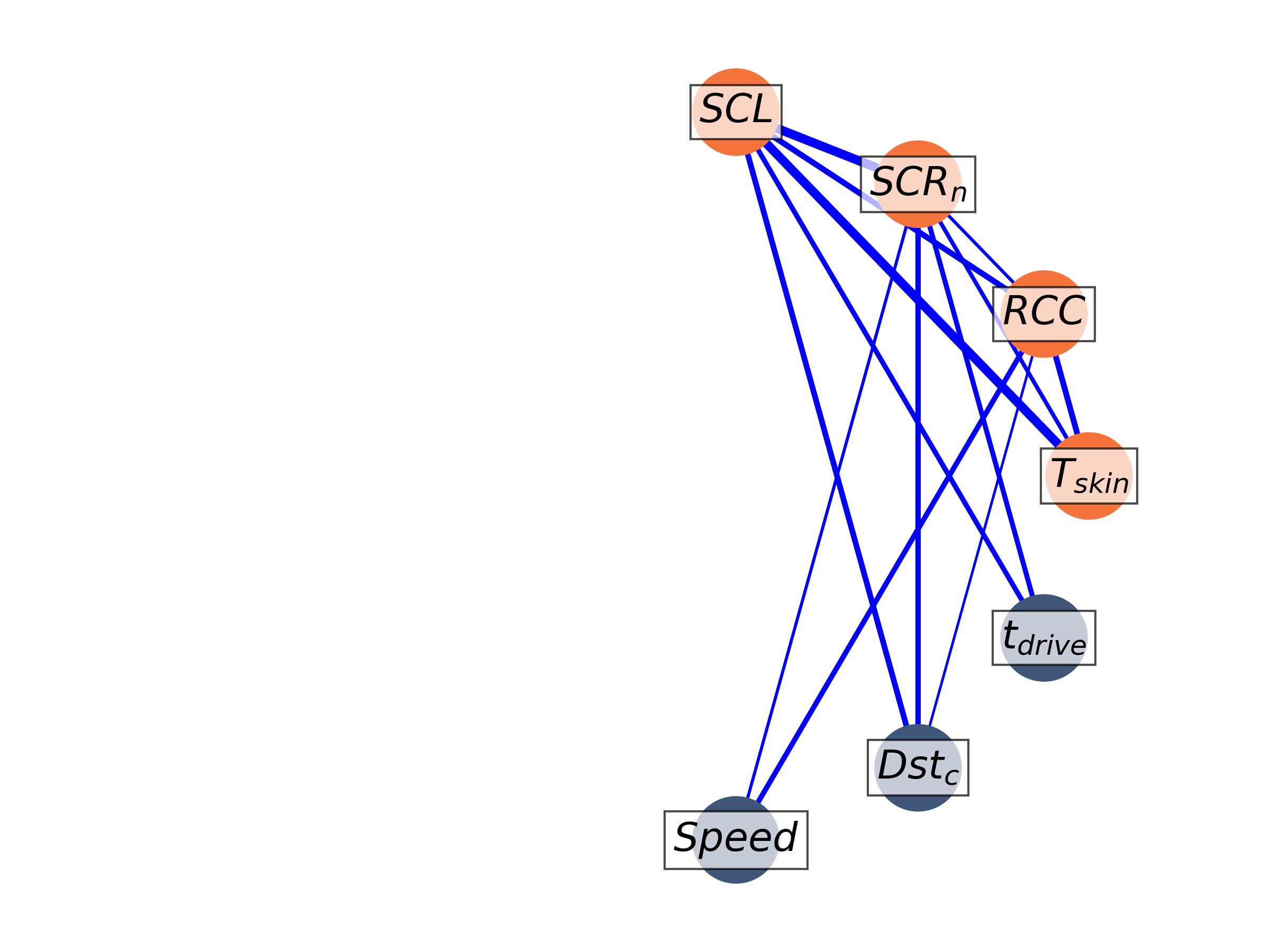} \label{fig:activity_phys}}
\subfloat[Demographics and physiological \\biomarkers]
{\includegraphics[width=0.33\textwidth]{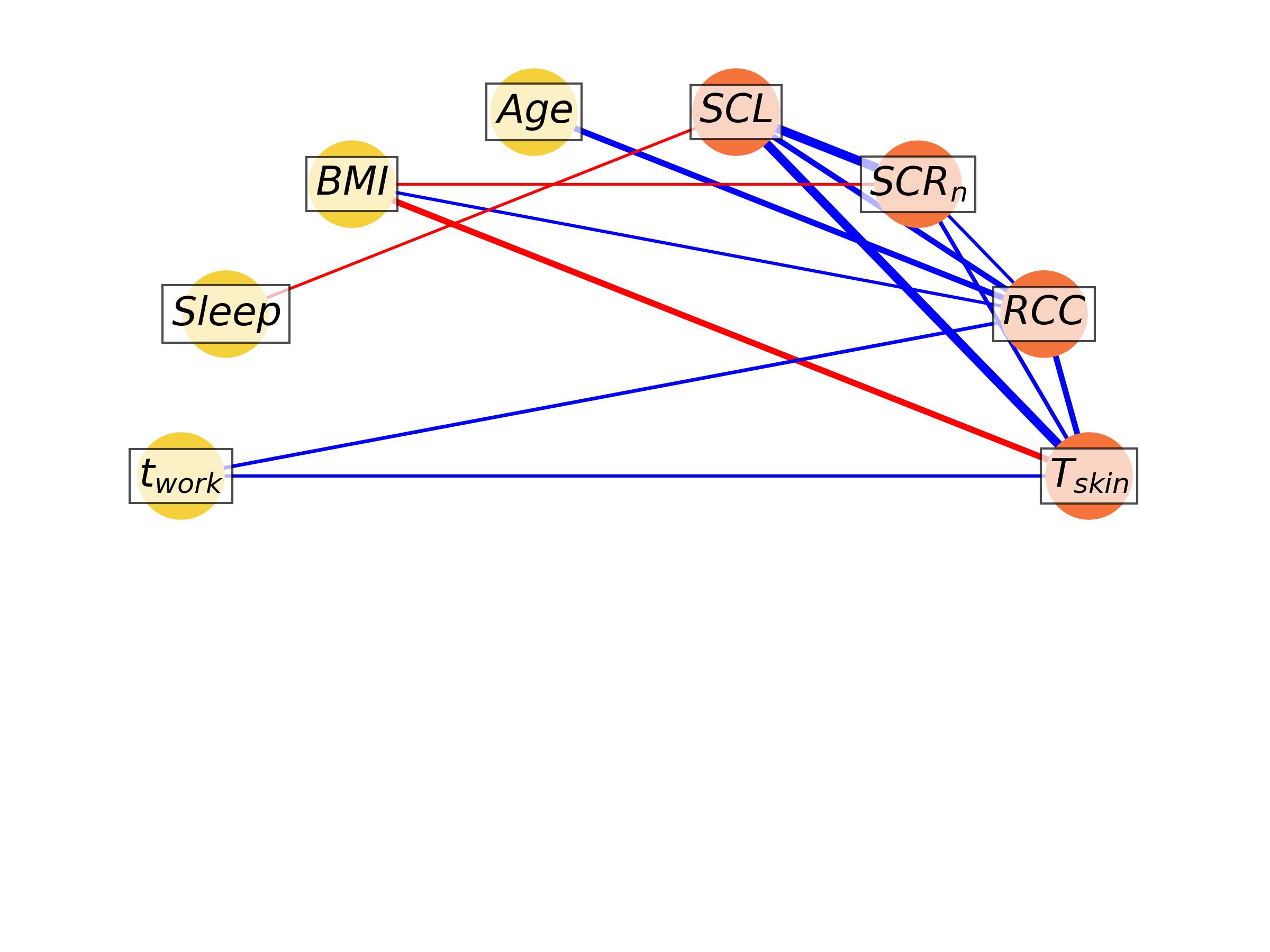} \label{fig:demog_phys}}
\subfloat[Weather and physiological \\biomarkers]
{\includegraphics[width=0.33\textwidth]{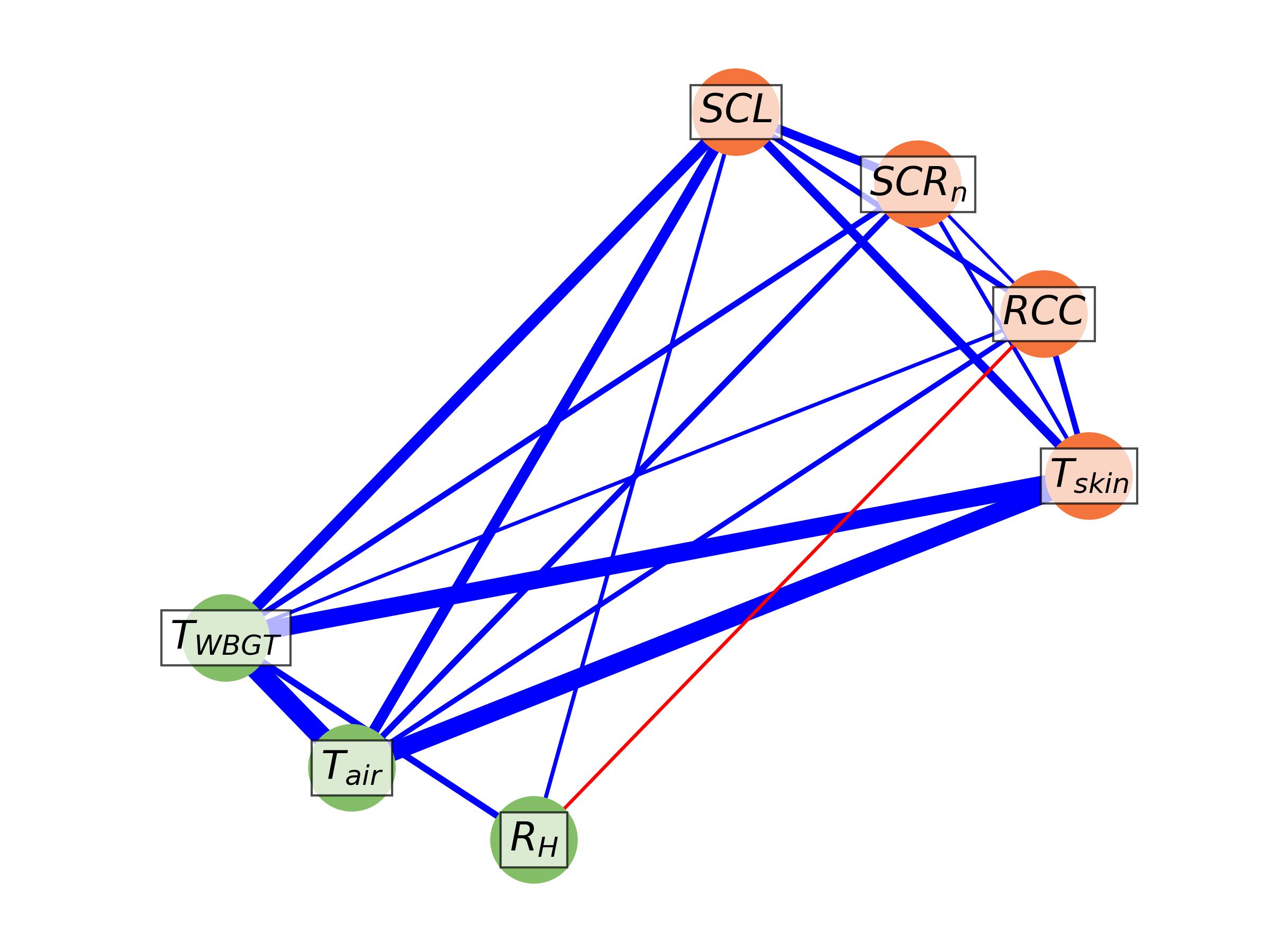} 
\label{fig:env_phys}}

\caption{Correlation network based on statistically significant Pearson correlation (Coefficient \textgreater= 0.1 and P $<$ 0.05) between two variables. Blue-colored and red-colored edges represent positive and negative correlations, respectively. 
}
\label{fig:correlation_netoworks}
\end{figure*}

\subsection{Results}
The resulting correlation matrix is presented in Table \ref{tab:correlation_matrix}, with statistically significant correlations indicated by asterisks. The abbreviation of each feature in the correlation network is illustrated in Table \ref{tab: features}.
Figure \ref{fig:correlation_netoworks} illustrates the corresponding correlation network, with blue and red edges representing positive and negative correlations, respectively. The width of edges reflects the strength of correlation.

Activity features such as speed ($Speed$), duration of driving ($t_{drive}$), and cumulative distance ($Dst_c$) exhibit strong correlations with most of the physiological variables, as illustrated in Figure \ref{fig:activity_phys}. Strong positive correlations suggest that those rickshaw pullers who drive with more intensive activity have their skin temperature tend to increase. Besides, more speed requires more cardiac cost and activates the sweat gland.

Demographic variables such as $Age$ show a positive correlation with $RCC$. This signifies that older rickshaw pullers need to put a greater percentage of their cardiac capacity in driving the rickshaw compared to younger ones. Besides, rickshaw pullers having more body mass index ($BMI$) sweat more (positive correlations with $SCL$) and have lower skin temperature (negative correlations with $T_{skin}$). 

Weather variables such as air temperature ($T_{air}$) and wet bulb globe temperature ($T_{WBGT}$) have a significant impact on rickshaw pullers' physiological biomarkers, as illustrated in Figure \ref{fig:env_phys}. These weather variables exhibit strong positive correlations with physiological biomarkers, particularly skin temperature ($T_{skin}$) and skin conductance level ($SCL$). 

We also observe correlations among the physiological variables. $T_{skin}$ is correlated with $SCL$ reflecting the thermoregulation and the correlation with $RCC$ signifies that increasing cardiac output is needed to regulate $T_{skin}$.

\begin{figure}[!tbp]
\centering

  \centering
    \includegraphics[width=\linewidth]{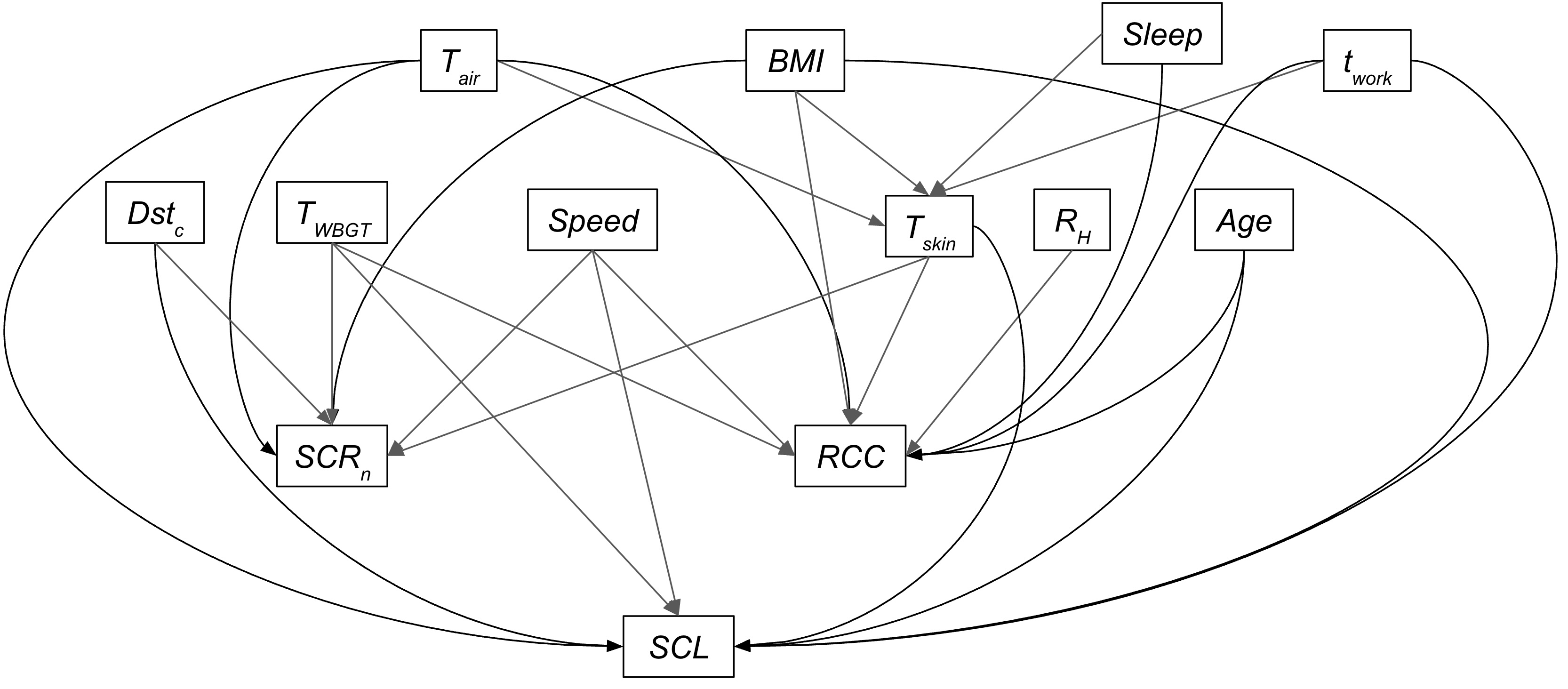}

\caption{The directed acyclic graph shows the Linear Gaussian Bayesian Network (LGBN) with the best validation score among other networks. It captures the directional dependencies among environmental, physiological, and demographic variables.
}
\label{fig:a_BBN_Network}
\end{figure}

\section{Regression Framework}
\textcolor{red}{As the feature space of our dataset is continuous (Table \ref{tab: features}), we implement a regression-based modeling framework. In the remaining subsections, we present this framework.}
\subsection{Method}
\textcolor{red}{We adopt a subject-wise train–validation–test protocol: 50 participants are reserved for training, 25 for validation, and 25 are held out for testing. A broad suite of regression models is evaluated. Baselines include linear regression, support vector regression, Random Forests, XGBoost, and Multi-Layer Perceptron. Recent table-focused models, such as TabNet and TabTransformer \cite{huang2020tabtransformer, arik2021tabnet}, are also assessed alongside Linear Gaussian Bayesian Network (LGBN).  }


\subsubsection{Training}

\textcolor{red}{We configure the Random Forest model with 300 decision trees of unbounded depth. For gradient boosting (XGBoost), we use 300 shallow trees with a maximum depth of 3 and a learning rate of 0.01. For the MLP, we tune hyperparameters on the validation set after the initial training, and then implement a single hidden layer with 64 units and tanh activation, optimized with Adam (learning rate = 0.001) for 500 iterations.}

\textcolor{red}{In the case of LGBN, multiple directed acyclic graphs (DAGs) are learned from the training data using structure learning algorithms, such as Hill-Climb Search, Greedy Equivalence Search (GES), and NoTears \cite{NEURIPS2018_e347c514}. Bayesian Information Criterion (BIC) score for Gaussian Bayesian Network structures is used for model selection in Hill-Climb Search and GES. BIC score rewards goodness of fit and penalizes model complexity. For NoTears, we experiment with multiple weight thresholds, which serve as penalization parameters that prune weaker edges from the learned DAGs. For different weight thresholds (e.g., 0.05, 0.1, and 0.2), a corresponding DAG structure is learned. The resulting candidate DAGs from each algorithm and weight thresholds are saved for validation. Experiments are performed using CausalNex and pgmpy library \cite{causalnex, pgmpy}.
In our implementation of Linear Gaussian Bayesian Networks (LGBNs), we do not impose explicit priors over the conditional variances of the Gaussian distributions. Instead, variance parameters are estimated directly from the training data using maximum likelihood estimation.

\subsubsection{Validation}
During validation, Linear Gaussian Bayesian Networks are estimated for each candidate DAG. These estimated models are then fitted using the training data to obtain the corresponding conditional probability distributions. Afterward, physiological variables ($T_{kin}$, $RCC$, $SCR_n$, $SCL$) are predicted for the validation data, and Normalized Mean Absolute Error (NMAE) is measured. The average NMAE across all physiological variables is considered an evaluation metric for model selection, which means the DAG that gives the least average NMAE is considered the best network structure. Afterward, this structure is fitted with both the training and validation data and tested against the held-out dataset.}

\textcolor{red}{Here is the validation performance of LGBN for the candidate DAG structures: Hill-Climb Search (NMAE = 0.1547), GES (NMAE = 0.1547), NoTears with weight thresholds of 0.05 (NMAE = 0.1532), 0.1 (NMAE = 0.1502), and 0.2 (NMAE = 0.1545). NoTears model, with weight threshold = 0.1, delivers the best fit and is therefore selected for test on held-out data. Figure \ref{fig:a_BBN_Network} illustrates the best-performing DAG structure. }

\begin{table}[!tbp]
\centering
\caption{Performance comparison of regression models across physiological biomarkers. MAE, NMAE, and correlation coefficient are reported for each model–variable combination. Bold values indicate the best performance for each metric within a target variable.}
\label{tab:results_regression}
\resizebox{\linewidth}{!}{

\begin{tabular}{|l|l|l|l|l|l|}
\hline
\begin{tabular}[c]{@{}l@{}}Target \\ Variable\end{tabular} & Model & \begin{tabular}[c]{@{}l@{}}Feature combination based \\ on mutual information index\end{tabular} & MAE & NMAE & \begin{tabular}[c]{@{}l@{}}Correlation \\ Coefficient\end{tabular} \\ \hline
\multirow{8}{*}{$T_{skin}$} & Linear & \multirow{5}{*}{\begin{tabular}[c]{@{}l@{}}$T_{WBGT}$, $T_{air}$, $R_H$, $BMI$, $Age$, \\ $t_{work}$, $t_{drive}$, $Dst_c$, $Sleep$, \\ $Speed$\end{tabular}} & 0.82 (range: 27.7–37.0, mean 33.2) & \textbf{0.079} & \textbf{0.824} \\ \cline{2-2} \cline{4-6} 
 & Support Vector Regressor &  & 1.14 (range: 27.7–37.0, mean 33.2) & 0.119 & 0.811 \\ \cline{2-2} \cline{4-6} 
 & Random Forest &  & 0.85 (range: 27.7–37.0, mean 33.2) & 0.089 & 0.814 \\ \cline{2-2} \cline{4-6} 
 & XGBoost &  & 0.85 (range: 27.7–37.0, mean 33.2) & 0.089 & 0.803 \\ \cline{2-2} \cline{4-6} 
 & MLP &  & 4.10 (range: 27.7–37.0, mean 33.2) & 0.430 & 0.220 \\ \cline{2-6} 
 & TabTransformer & \multirow{3}{*}{-} & 3.03 (range: 27.7–37.0, mean 33.2) & 0.325 & 0.483 \\ \cline{2-2} \cline{4-6} 
 & TabNet &  & 1.53 (range: 27.7–37.0, mean 33.2) & 0.166 & 0.246 \\ \cline{2-2} \cline{4-6} 
 & \begin{tabular}[c]{@{}l@{}}Linear Gaussian Bayesian \\ Network {[}NoTears{]}\end{tabular} &  & 0.86 (range: 27.7–37.0, mean 33.2) & 0.093 & 0.798 \\ \hline
\multirow{8}{*}{$RCC$} & Linear & \multirow{5}{*}{\begin{tabular}[c]{@{}l@{}}$R_H$, $T_{WBGT}$, $T_{air}$, $BMI$, $Age$\\ $t_{drive}$, $Dst_c$, $t_{work}$, $Speed$, \\ $Sleep$\end{tabular}} & 9.40 (range: 6.4–63.0, mean 27.5) & 0.176 & 0.372 \\ \cline{2-2} \cline{4-6} 
 & Support Vector Regressor &  & 10.39 (range: 6.4–63.0, mean 27.5) & 0.195 & 0.382 \\ \cline{2-2} \cline{4-6} 
 & Random Forest &  & 9.13 (range: 6.4–63.0, mean 27.5) & 0.171 & 0.340 \\ \cline{2-2} \cline{4-6} 
 & XGBoost &  & 9.23 (range: 6.4–63.0, mean 27.5) & 0.173 & 0.340 \\ \cline{2-2} \cline{4-6} 
 & MLP &  & 9.02 (range: 6.4–63.0, mean 27.5) & \textbf{0.169} & 0.385 \\ \cline{2-6} 
 & TabTransformer & \multirow{3}{*}{-} & 11.77 (range: 6.4–63.0, mean 27.5) & 0.208 & 0.449 \\ \cline{2-2} \cline{4-6} 
 & TabNet &  & 10.56 (range: 6.4–63.0, mean 27.5) & 0.187 & 0.296 \\ \cline{2-2} \cline{4-6} 
 & \begin{tabular}[c]{@{}l@{}}Linear Gaussian Bayesian \\ Network {[}NoTears{]}\end{tabular} &  & 10.37 (range: 6.4–63.0, mean 27.5) & 0.183 & \textbf{0.467} \\ \hline
\multirow{8}{*}{$SCR_n$} & Linear & \multirow{5}{*}{\begin{tabular}[c]{@{}l@{}}$Dst_c$, $T_{WBGT}$, $t_{drive}$, $T_{air}$, \\ $R_H$, $Speed$, $BMI$, $Age$, $t_{work}$, \\ $Sleep$\end{tabular}} & 8.23 (range: 0.0–49.0, mean 14.8) & 0.175 & 0.573 \\ \cline{2-2} \cline{4-6} 
 & Support Vector Regressor &  & 7.96 (range: 0.0–49.0, mean 14.8) & 0.169 & 0.593 \\ \cline{2-2} \cline{4-6} 
 & Random Forest &  & 8.28 (range: 0.0–49.0, mean 14.8) & 0.176 & 0.580 \\ \cline{2-2} \cline{4-6} 
 & XGBoost &  & 7.77 (range: 0.0–49.0, mean 14.8) & 0.165 & 0.663 \\ \cline{2-2} \cline{4-6} 
 & MLP &  & 10.06 (range: 0.0–49.0, mean 14.8) & 0.214 & 0.406 \\ \cline{2-6} 
 & TabTransformer & \multirow{3}{*}{-} & 8.96 (range: 0.0–49.0, mean 14.8) & 0.183 & 0.472 \\ \cline{2-2} \cline{4-6} 
 & TabNet &  & 8.52 (range: 0.0–49.0, mean 14.8) & 0.174 & 0.529 \\ \cline{2-2} \cline{4-6} 
 & \begin{tabular}[c]{@{}l@{}}Linear Gaussian Bayesian \\ Network {[}NoTears{]}\end{tabular} &  & 7.43 (range: 0.0–49.0, mean 14.8) & \textbf{0.152} & \textbf{0.653} \\ \hline
\multirow{8}{*}{$SCL$} & Linear & \multirow{5}{*}{\begin{tabular}[c]{@{}l@{}}$T_{WBGT}$, $T_{air}$, $BMI$, $Age$, $R_H$, \\ $t_{work}$, $Sleep$, $t_{drive}$, \\ $Dst_c$, $Speed$\end{tabular}} & 13.60 (range: 0.0–76.2, mean 22.8) & 0.119 & 0.543 \\ \cline{2-2} \cline{4-6} 
 & Support Vector Regressor &  & 14.52 (range: 0.0–76.2, mean 22.8) & 0.127 & 0.570 \\ \cline{2-2} \cline{4-6} 
 & Random Forest &  & 13.29 (range: 0.0–76.2, mean 22.8) & \textbf{0.116} & 0.541 \\ \cline{2-2} \cline{4-6} 
 & XGBoost &  & 13.42 (range: 0.0–76.2, mean 22.8) & 0.117 & 0.553 \\ \cline{2-2} \cline{4-6} 
 & MLP &  & 15.73 (range: 0.0–76.2, mean 22.8) & 0.138 & 0.428 \\ \cline{2-6} 
 & TabTransformer & \multirow{3}{*}{-} & 12.54 (range: 0.0–76.2, mean 22.8) & 0.165 & 0.644 \\ \cline{2-2} \cline{4-6} 
 & TabNet &  & 16.56 (range: 0.0–76.2, mean 22.8) & 0.217 & 0.201 \\ \cline{2-2} \cline{4-6} 
 & \begin{tabular}[c]{@{}l@{}}Linear Gaussian Bayesian \\ Network {[}NoTears{]}\end{tabular} &  & 16.04 (range: 0.0–76.2, mean 22.8) & 0.210 & \textbf{0.674} \\ \hline
\end{tabular}

}
\end{table}

\subsection{Results}

Table \ref{tab:results_regression} presents the performance of the validated models on held-out test data. Experimental results show that model performance varies by physiological variables. 

\subsubsection{Quantitative Model Performance}
For skin temperature ($T_{skin}$), the linear regressor achieves the best performance, with a Mean Absolute Error (MAE) of 0.824, a Normalized MAE (NMAE) of 0.079, and a correlation coefficient of 0.824. For relative cardiac cost ($RCC$), the multilayer perceptron (MLP) yields the lowest NMAE (0.169), while the Linear Gaussian Bayesian Network (NoTears) achieves the highest correlation coefficient (0.467). For skin conductance response ($SCR_n$), the Linear Gaussian Bayesian Network consistently outperforms other models, with an NMAE of 0.152 and a correlation coefficient of 0.653. For skin conductance level ($SCL$), the Random Forest regressor records the lowest NMAE (0.116), whereas the Linear Gaussian Bayesian Network again provides the highest correlation coefficient (0.674). The tabular deep learning models, TabTransformer and TabNet \cite{huang2020tabtransformer, arik2021tabnet} seem to be showing moderate performance across all target variables, as detailed in Table \ref{tab:results_regression}.

\subsubsection{Scatter Plot Evaluation}
Scatter plots of true versus predicted values for the Linear Gaussian Bayesian Network are provided in Supplementary Figures 3 and 4. These plots show that the model predictions for skin temperature ($T_{skin}$) align closely with the ideal line, indicating strong agreement with ground-truth values. In contrast, predictions for RCC, SCL, and $SCR_n$ exhibit greater dispersion, reflecting higher variability in model performance across biomarkers. Nevertheless, clustering of points along the diagonal line suggests that the model captures key trends in the data, even when residual errors remain moderate.

\subsubsection{Summary and Model Selection}
Overall, the findings highlight that model performance is biomarker-specific: Linear Regression provides the best estimates for skin temperature, MLP and the Linear Gaussian Bayesian Network (LGBN) are most effective for relative cardiac cost, LGBN outperforms others for skin conductance response, and both Random Forest and LGBN achieve superior performance for skin conductance level. Among these, we select the LGBN for integration with climate model outputs and subsequent survivability analysis, due to its interpretability and ability to represent causal relationships within the data.

\section{Climate Model-Based Forecast}
To project how future climate conditions may influence physiological stress in rickshaw pullers, we integrate climate model outputs into our regression framework. In this section, we describe how climate projections are obtained and incorporated into the forecasting pipeline.

\subsection{Method}
In Figure \ref{fig:ml_and_climate}, we demonstrate how climate modeling can be used to forecast physiological biomarkers in the future.
Considering different emission scenarios, we aim to forecast the changes in weather. Our approach involves integrating projected changes in temperature and humidity from climate models into the regression-based model to anticipate future physiological responses. 

\textcolor{red}{Since climate models span a wide range of climate sensitivities, which is a key uncertainty in climate modeling, we choose to forecast the climate of the region of Bangladesh from 2026 to 2100 for 18 CMIP6 global climate models (GCM). 
The models include: ACCESS-CM2, AWI-CM-1-1-MR, BCC-CSM2-MR, CAMS-CSM1-0, CESM2-WACCM, CMCC-CM2-SR5, CanESM5, EC-Earth3-Veg-LR, FGOALS-f3-L, FGOALS-g3, GFDL-ESM4, IITM-ESM, INM-CM4-8, INM-CM5-0, IPSL-CM6A-LR, KACE-1-0-G, MIROC6, MPI-ESM1-2-HR, MPI-ESM1-2-LR, MRI-ESM2-0, NorESM2-LM, NorESM2-MM, and TaiESM1 \cite{gmd-9-1937-2016}.}
Though these models are designed to capture global climate, they have a spatial resolution of at least 1$^\circ$ and a temporal resolution of each month of a year covering 1850-2100. We interpolate the surface temperature and relative humidity for the region of Bangladesh (Latitude: [20.7505235, 26.6325753] and Longitude: [88.0363086, 92.6820672]) from the climate models. 

\textcolor{red}{Apart from specifying resolution, we select four Shared Socioeconomic Pathways (SSP126, SSP245, SSP370, and SSP585). Each one of the SSPs demonstrates a specific greenhouse gas emission in the future, where lower SSPs represent less emission and higher ones represent more emission. SSP1 (Sustainability - Taking the green road ) assumes the world is shifting gradually towards a more sustainable path, respecting the environmental boundary. SSP2 (Middle of the road) assumes a world where social, economic, and technological trends do not shift significantly with respect to historical trends. SSP3 (Regional rivalry - A rocky road) assumes a competitive scenario over the world where countries focus on regional issues in achieving energy and food security. SSP4 (Inequality - A road divided) assumes unequal investments in human capital, leading to increasing inequalities both across and within countries. SSP5 (Fossil-fueled development - Taking the highway) assumes a world where the push for economic and social development is coupled with the demand for abundant fossil fuel resources \cite{RIAHI2017153, gmd-9-1937-2016}. Among the listed SSPs, four scenarios (SSP126, SSP245, SSP370, and SSP585) are designated as Tier 1 priority scenarios by the CMIP6 community, meaning they are strongly recommended for experimentation due to their scientific and policy relevance. Among these, we emphasize SSP245 because it represents the `middle-of-the-road' socioeconomic pathway, closest to today’s aggregate emission trajectory.}

Going forward, based on the climate model outputs, we feed the change in $T_{air}$, $R_H$, and $SR$ to our prepared dataset. Specifically, we modify weather variables (e.g., $T_{air}$, $R_H$, and $T_{WBGT}$) while keeping activity (e.g., $Speed$, $Dst_c$, and $t_{drive}$) unchanged. This ensures the conduction of a simulation where only the climate has changed, and rickshaw pullers have to make the same trip as before. Finally, we run the regression model on the updated dataset and infer the physiological variables.

\begin{figure*}[!tbp]
\centering
\subfloat[Surface temperature]
{\includegraphics[width=0.49\textwidth]{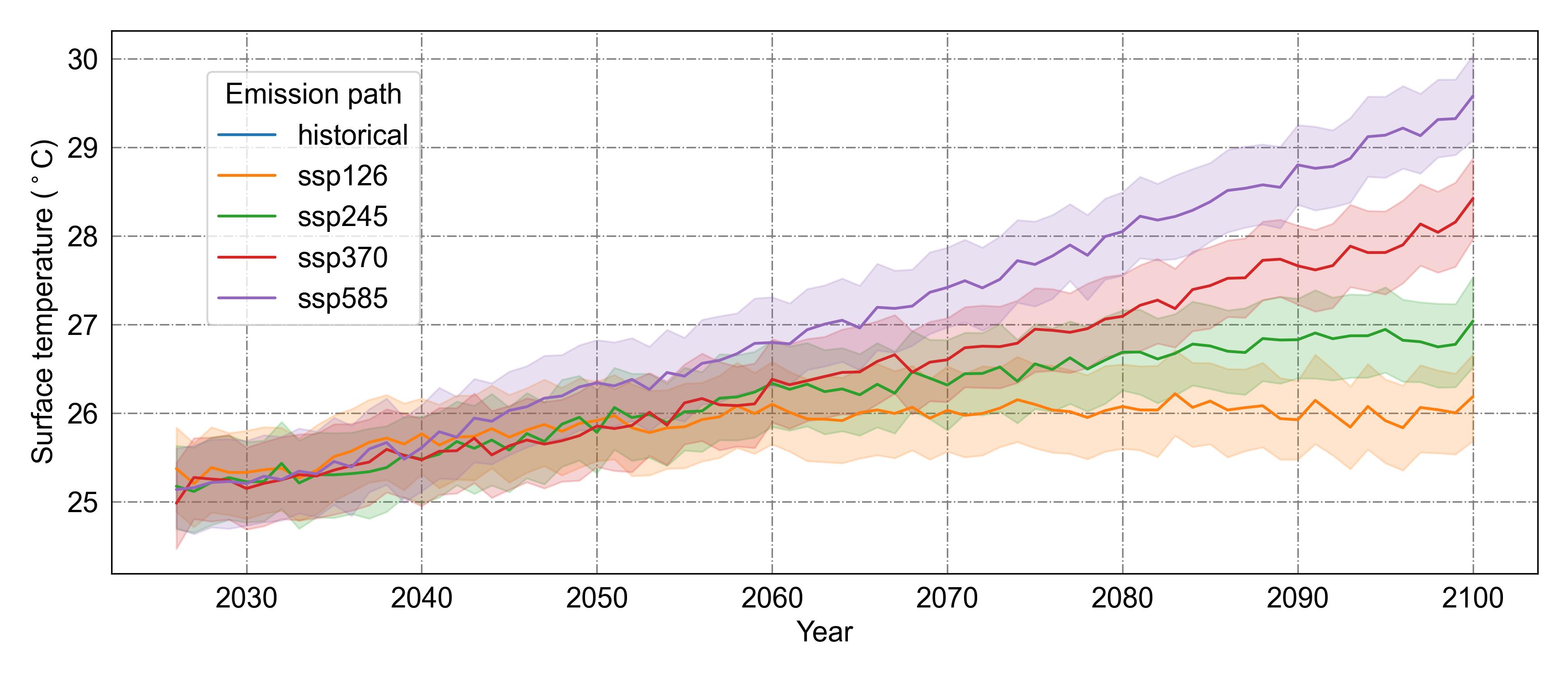} \label{fig:all_tas_with_historical}}
\hfill
\subfloat[Relative humidity]
{\includegraphics[width=0.49\textwidth]{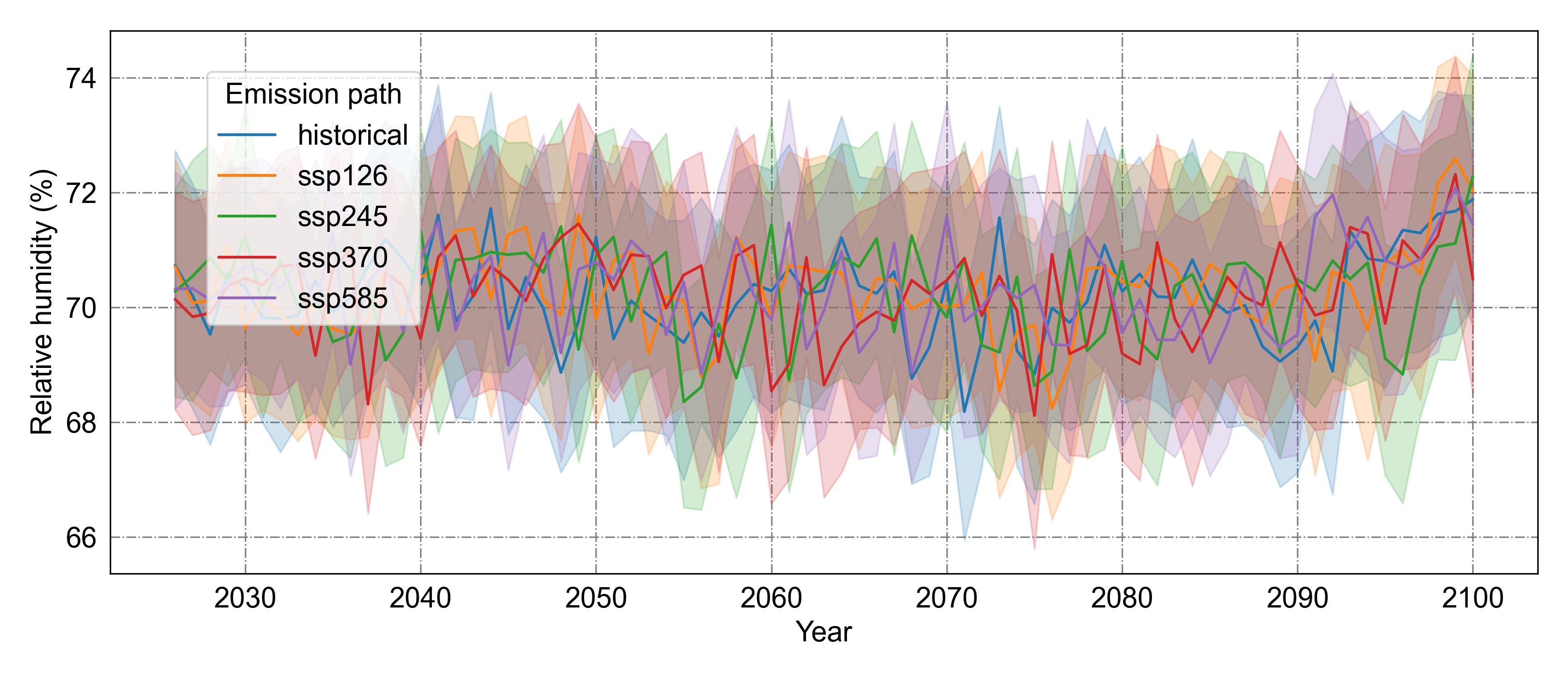} \label{fig:all_hum_with_historical}}
\hfill
\subfloat[Solar radiation]
{\includegraphics[width=0.49\textwidth]{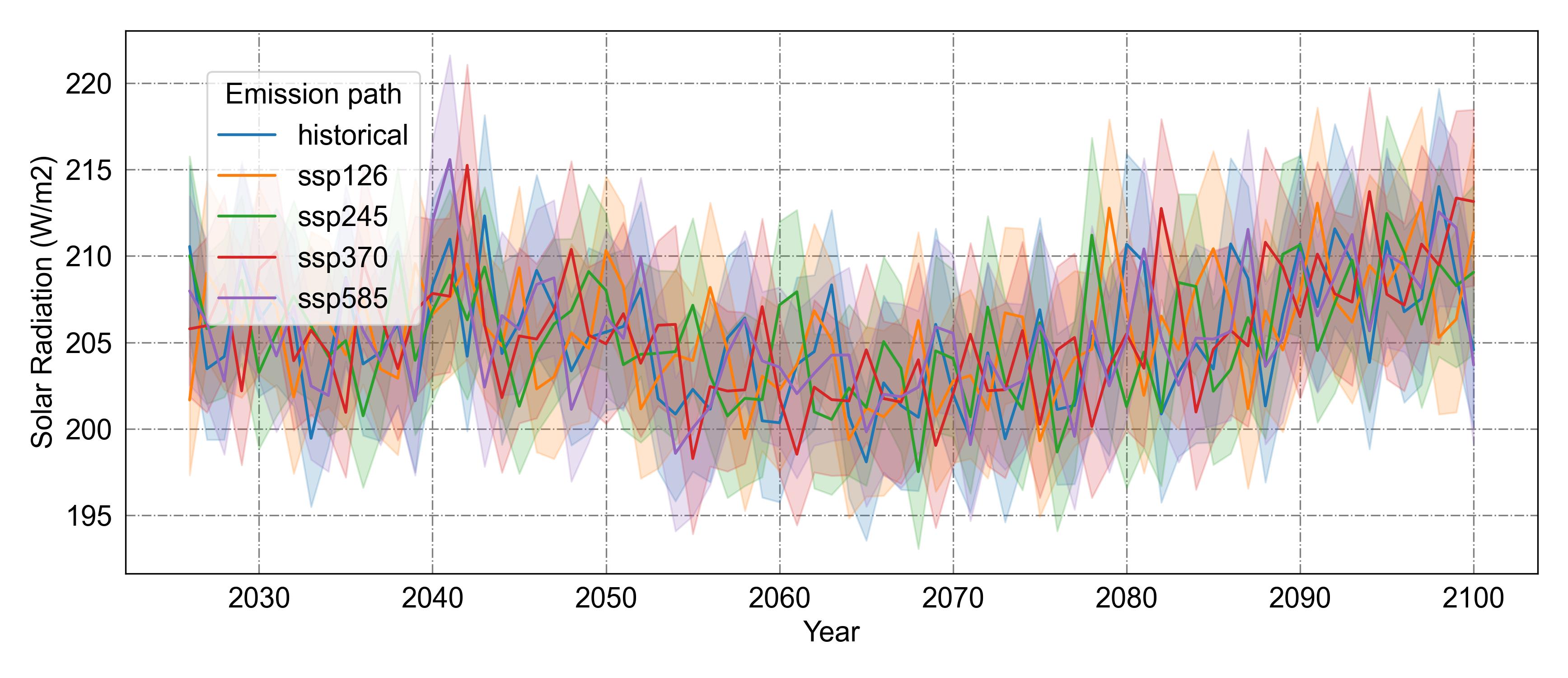} \label{fig:all_solar_radiation_with_historical}}
\hfill
\subfloat[Wet bulb globe temperature]
{\includegraphics[width=0.49\textwidth]{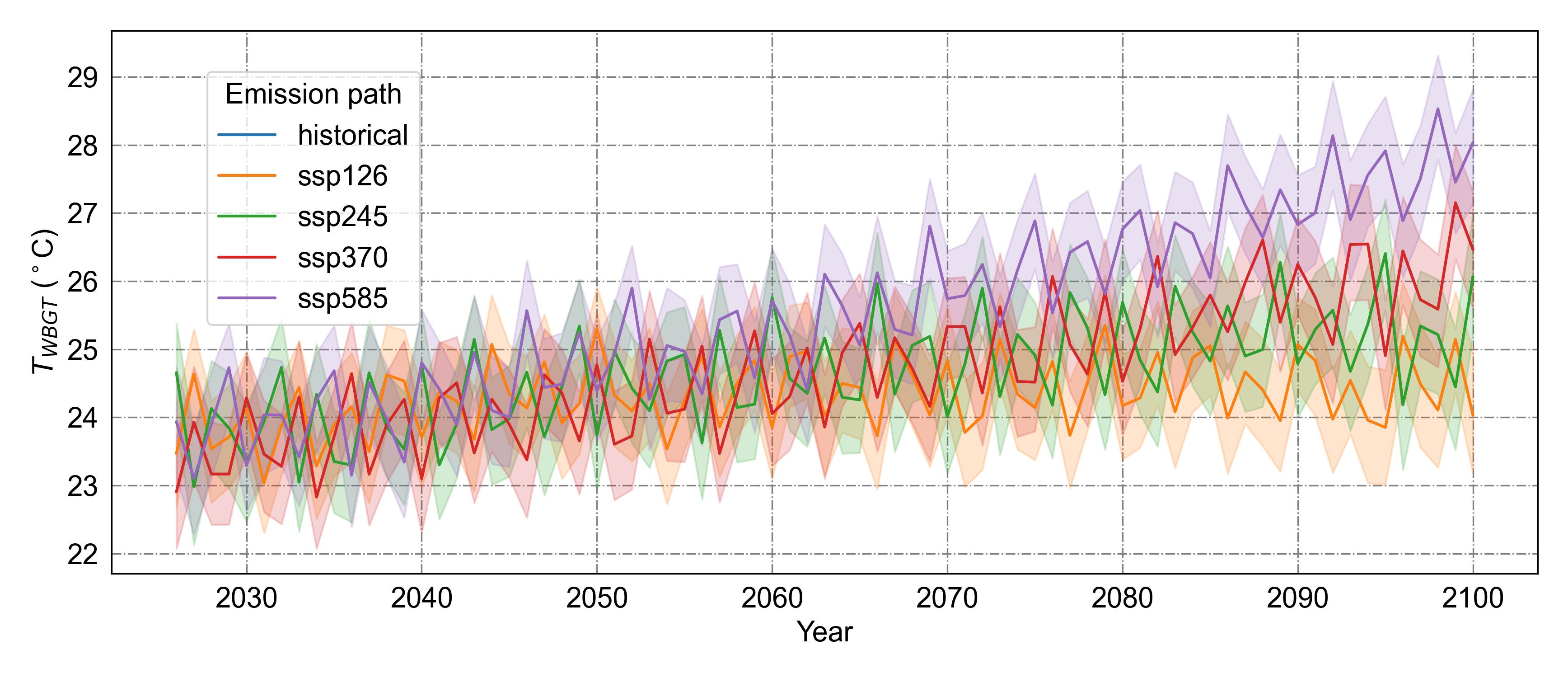} \label{fig:all_wbgt_with_historical}}
\caption{Climate model-based prediction (2023-2100) of surface temperature, relative humidity, solar radiation, and wet bulb globe temperature in Bangladesh based on 18 CMIP6 climate models.	}
\label{fig:climate_model_inference_environment}

\end{figure*}

\subsection{Results}

In Figure \ref{fig:climate_model_inference_environment}, we present the historical and forecasted surface temperature, relative humidity, solar radiation, and wet bulb globe temperature (WBGT) for Bangladesh. The shaded areas represent the 95\% confidence intervals (CI) across different climate models. Surface temperature is projected to increase by at least 1$^\circ$C by 2100 even under the lowest emission scenario (SSP126), while relative humidity remains relatively stable with only a minor decreasing trend.

Based on the climate model outputs, to evaluate the reliability of each physiological variable for long-term forecasting, we conducted an uncertainty analysis of the LGBN predictions for skin temperature ($T_{skin}$), relative cardiac cost ($RCC$), skin conductance response ($SCR_n$), and skin conductance level ($SCL$). A detailed description of the bootstrapping-based Monte Carlo procedure and full results are provided in Supplementary Section 1.1 and Supplementary Figure 1. In summary, $T_{skin}$ exhibits by far the lowest prediction uncertainty, with extremely narrow confidence intervals and minimal variance across bootstrap samples. $RCC$ and $SCR_n$ show moderate uncertainty, while $SCL$ demonstrates substantial variability, indicating limited predictive stability. These results indicate that although multiple biomarkers capture aspects of thermal strain, their robustness as predictive outcomes varies considerably. Skin temperature emerges as the most stable, consistent, and environmentally responsive biomarker, making it the most reliable candidate for climate-integrated survivability analysis.

\section{Survivability Analysis}

To understand how future climate conditions may threaten the physiological safety of rickshaw pullers, we translate environmental and physiological projections into survivability risk. In this section, we analyze survivability using established thresholds and model future exposure patterns.

\subsection{Method}
\label{survivability: method}
We first review survivability metrics and thresholds proposed in the literature that link environmental and physiological stress, focusing on wet bulb globe temperature ($T_{WBGT}$) and skin temperature ($T_{skin}$) as two widely accepted indicators of heat-related risk \cite{Sherwood2010, BUDD200820, Vanos2023}. These metrics, along with their corresponding survivability thresholds and physiological consequences, are summarized in Table \ref{tab:survivavility_thresholds}. Using these established thresholds, we assess the survivability of rickshaw pullers under both current (2023–2025) and projected future climate conditions.

We focus on $T_{WBGT}$ and $T_{skin}$ because they capture complementary dimensions of heat stress: $T_{WBGT}$ reflects external environmental load, while $T_{skin}$ represents the body’s internal thermophysiological response. Although skin temperature is influenced by external factors (e.g., ambient temperature, clothing), it directly affects core body temperature, as the skin must remain cooler than the core to enable dissipation of metabolic heat \cite{Sherwood2010}. Prolonged skin temperatures above 35 °C signal elevated core temperature, and values reaching 37–38 °C can drive core temperature toward potentially lethal levels (42–43 °C), even in physically fit individuals. Therefore, $T_{skin}$ serves as a critical early-warning biomarker of survivability under extreme heat exposure.

\textcolor{red}{Going forward, as climate model projections inherently involve uncertainty, stemming from the ensemble of over 18 CMIP6 models, we conduct an uncertainty analysis encompassing both the Linear Gaussian Bayesian Network (LGBN) and the climate model outputs. The methodology and implementation details of this uncertainty quantification are presented in Section 1 of the supplementary document. Based on this analysis, we report the $T_{WBGT}$ and $T_{skin}$ exposure metrics alongside their standard deviations from the mean estimates, thereby communicating the predictive uncertainty associated with future physiological risk projections.}

\begin{table}[!tbp]
\centering
\caption{Survivability variables and corresponding thresholds with heat stress levels \cite{Sherwood2010, BUDD200820, Vanos2023}}
\label{tab:survivavility_thresholds}
\begin{tabular}{|c|c|l|l|}
\hline
Type & Variable & \multicolumn{1}{c|}{Threshold ($^\circ$C)} & \multicolumn{1}{c|}{Physiological condition} \\ \hline
\multirow{5}{*}{Weather} & \multirow{5}{*}{$T_{WBGT}$} & 25.6 - 27.8 & Good conditions or no stress \\ \cline{3-4} 
 &  & 27.8 - 29.4 & \begin{tabular}[c]{@{}l@{}}Less than ideal conditions or \\ mild risk\end{tabular} \\ \cline{3-4} 
 &  & 29.4 - 31.1 & \begin{tabular}[c]{@{}l@{}}Moderate risk of heat-related\\ illness\end{tabular} \\ \cline{3-4} 
 &  & 31.1 - 32.2 & High risk of heat-related illness \\ \cline{3-4} 
 &  & \textgreater 32.2 & Extreme conditions \\ \hline
\multirow{2}{*}{\begin{tabular}[c]{@{}l@{}}Physiological \\ biomarker\end{tabular}} & \multirow{2}{*}{$T_{skin}$} & 35 & Normal condition \\ \cline{3-4} 
 &  & \textgreater 35 & \begin{tabular}[c]{@{}l@{}}Prolonged exposure can cause an \\ increase in core body temperature\end{tabular} \\ \hline
\end{tabular}
\end{table}

\begin{table}[!tbp]
\caption{\textcolor{red}{ Demonstration of exposure to extreme heat in present and future (SSP245) based on weather ($T_{WBGT}$) and physiological biomarker ($T_{skin}$). A total of 100 subjects participated in our study. To assess exposure, the total number of subjects exposed and on average, how long the exposure has been sustained are also measured. Exposure categories are based on studies done by Sherwood et al., and Vanos et al., which are discussed broadly in Section \ref{survivability: method} \cite{Sherwood2010, BUDD200820, Vanos2023}.} }
\label{tab:compare_surviabailiy_present_future}
\resizebox{\linewidth}{!}{

\begin{tabular}{|l|l|l|l|l|l|l|l|l|l|l|}
\hline
Exposure category & Metrics & 
\begin{tabular}[c]{@{}l@{}}2023–\\2025\end{tabular} & 
\begin{tabular}[c]{@{}l@{}}2026–\\2030\end{tabular} & 
\begin{tabular}[c]{@{}l@{}}2031–\\2040\end{tabular} & 
\begin{tabular}[c]{@{}l@{}}2041–\\2050\end{tabular} & 
\begin{tabular}[c]{@{}l@{}}2051–\\2060\end{tabular} & 
\begin{tabular}[c]{@{}l@{}}2061–\\2070\end{tabular} & 
\begin{tabular}[c]{@{}l@{}}2071–\\2080\end{tabular} & 
\begin{tabular}[c]{@{}l@{}}2081–\\2090\end{tabular} & 
\begin{tabular}[c]{@{}l@{}}2091–\\2100\end{tabular} \\ \hline

\multirow{3}{*}{\begin{tabular}[c]{@{}l@{}}$T_{WBGT}$ \textgreater{}31.1: High \\ risk of heat-related \\ illness\\ or\\ Extreme conditions\end{tabular}} & \begin{tabular}[c]{@{}l@{}}Participants under \\ exposure (\%)\end{tabular} & 32 & 37 $\pm$17 & 39 $\pm$16 & 43 $\pm$16 & 46 $\pm$15 & 49 $\pm$14 & 50 $\pm$14 & 51 $\pm$15 & 53 $\pm$15 \\ \cline{2-11} 
 & \begin{tabular}[c]{@{}l@{}}Exposure duration \\ (minutes)\end{tabular} & 10 & 11.9 $\pm$2.2 & 12.4 $\pm$2.2 & 13 $\pm$2.2 & 13.5 $\pm$2 & 13.8 $\pm$2 & 13.8 $\pm$2 & 14.2 $\pm$1.8 & 14.2 $\pm$2 \\ \cline{2-11} 
 & \begin{tabular}[c]{@{}l@{}}Exposure duration \\ (\% of trip length)\end{tabular} & 58 & 68 $\pm$12 & 70 $\pm$11 & 73 $\pm$11 & 75 $\pm$10 & 77 $\pm$10 & 77 $\pm$9 & 79 $\pm$9 & 79 $\pm$9 \\ \hline
\multirow{3}{*}{\begin{tabular}[c]{@{}l@{}}$T_{skin}$ \textgreater 35:\\ Prolonged exposure\\ can cause increase\\ in core body\\ temperature\end{tabular}} & \begin{tabular}[c]{@{}l@{}}Participants under \\ exposure (\%)\end{tabular} & 20 & 28 $\pm$18 & 29 $\pm$18 & 33 $\pm$17 & 36 $\pm$17 & 38 $\pm$16 & 39 $\pm$16 & 41 $\pm$16 & 42 $\pm$16 \\ \cline{2-11} 
 & \begin{tabular}[c]{@{}l@{}}Exposure duration \\ (minutes)\end{tabular} & 11 & 11.4 $\pm$2 & 11.6 $\pm$2 & 12.2 $\pm$2 & 12.6 $\pm$2. & 12.9 $\pm$2 & 13.1 $\pm$2 & 13.1 $\pm$2 & 13.4 $\pm$2 \\ \cline{2-11} 
 & \begin{tabular}[c]{@{}l@{}}Exposure duration \\ (\% of trip length)\end{tabular} & 64 & 62 $\pm$12 & 63 $\pm$12 & 67 $\pm$11 & 69 $\pm$11 & 72 $\pm$10 & 71 $\pm$10 & 72 $\pm$10 & 73 $\pm$10 \\ \hline
\end{tabular}

}
\end{table}
\subsection{Results}




In Table \ref{tab:compare_surviabailiy_present_future}, we demonstrate statistics on the present (2023-2024) and future physiological condition of rickshaw pullers based on skin temperature and wet bulb globe temperature. These statistics are derived from the data collected from 100 participants.
As the statistics are derived from data collected throughout the year, covering multiple seasons, months, and times of day, they provide a comparative overview across years and decades. Focusing on a single season (e.g., summer, monsoon, or winter) would yield substantially higher or lower vulnerability estimates than these year-round averages.

Based on the statistics, we can see, in the present scenario (2023-2025), when the study is conducted, 32\%  of participants are exposed to `High risk of heat-related illness’ or ‘Extreme conditions' ($T_{WBGT}$ > 31.1 $^\circ$C). The percentage can increase to 37\% $\pm$17 by 2026-2030, and it can be up to 53\% $\pm$15 over the decade of 2091-2100. 
Besides, in 2023-2025, the average length of exposure is 10 minutes, which is 58\% of the total duration of the trip \footnote{Average length of trip for our participants is 18.9 minutes}. As rickshaw pullers in our study work 10 hours a day on average, such exposure can lead to a significant amount of time they are exposed to extreme heat. By the end of this decade, it can reach up to 14.2$\pm$2 minutes of exposure (79\%$\pm$9 of the trip duration on average). This can significantly impact rickshaw pullers' occupational survivability.

Based on $T_{skin}$, in the present scenario (2023-2025), 20\% of participants are exposed to `$T_{skin}$ > 35$^\circ$C', which can increase up to 28\%$\pm18$ in 2026-2030 and up to 42\%$\pm16$ by the end of the century. On average, rickshaw pullers currently experience 11 minutes of heat exposure per trip (64\% of trip duration) in the present climate (2023–2025). This exposure is projected to increase slightly to 11.4$\pm$2 minutes (62\% $\pm$12) by 2026–2030, and to rise substantially to 13.4$\pm$2 minutes (73\% $\pm$10) by the end of this century.

\section{Interview of Rickshaw Pullers}

We recruited 12 rickshaw pullers with at least 10 years of experience driving rickshaws in Dhaka to capture their knowledge, perceptions, and experiences related to climate change and rickshaw driving. These insights provide important context for our collected weather and physiological data.

\subsection{Questionnaire Preparation and Interview Management}
We prepared the questionnaire in the local language, Bangla (Bengali), which is the first language of both the author who developed it and the rickshaw pullers who participated. Two of the authors interviewed the rickshaw pullers upon receiving informed consent to record the audio. Rickshaw pullers were asked about climate change, cardiac strain, perceived temperature, and the amount of sweating. The interviews last for 20-30 minutes. They were also asked about any possible change in those variables based on their experience of driving. The English translation of the questionnaire is provided as a supplementary document. 

The recorded audios of the interviews are transcribed into Bangla with the help of Google Speech-to-Text API. Transcripts are then translated back into English for further analysis. Three authors validate this process by listening to each audio file. Among those three authors, two are involved in conducting the interview. 
Afterward, the same three authors perform Thematic Analysis on the translated English version of the interviews. In doing the analysis, we follow the six steps mentioned in the book `Thematic Analysis' written by Virginia Braun and Victoria Clarke \cite{braun2012thematic}. 


\subsection{Coding and Theme Preparation}

For each interview participant, we begin by familiarizing ourselves with the data through repeated listening to the audio files and careful reading of the transcripts. Guided by our research questions, we identify codes that capture associations between sensor-based physiological variables and the rickshaw pullers’ professional experiences. In the first stage, we apply semantic coding, i.e., codes that directly reflect the rickshaw pullers’ observations, ensuring that our analysis remains grounded in their accounts. Wherever possible, the codes use the participants’ own words or closely related expressions, avoiding researcher-imposed concepts. We also ensure that codes are brief and concise, serving as quick labels to capture key ideas. Importantly, we generate codes only for transcript segments relevant to our research questions. During the coding process, each identified segment is highlighted in the transcript, and the corresponding code is recorded in a spreadsheet. After completing the first pass of all transcripts, we conduct a second review to refine existing codes and capture any new ones that emerge. For theme generation, we examine patterns of similarity across codes to extract distinctive and representative themes. In the following subsection, we present the themes that are identified in alignment with our research questions.


\subsection{ \textcolor{red}{Heat Intensity and Thermal Stress}}
\textcolor{red}{Thermal stress due to exposure in outdoor working conditions is an important theme based on our analysis. Particularly, the scarcity of shaded streets, narrow pathways, and urban structures has made it difficult for rickshaw pullers to have a comfortable working condition. Therefore, even though winter tends to be cooler, the streets of Dhaka remain hot throughout the year. The remaining seasons are hotter, particularly worse during the monsoon. As stated,}

\begin{quote}
\textcolor{red}{\textit{`While heat persists year around, what comes during the monsoon season, after Ashar (a Bangla month) is unlike anything I've experienced before. This particular heat feels unprecedented.'}}
\begin{flushright}
\textcolor{red}{\textit{--- A 52-year-old rickshaw puller, experience driving for 25 years}}
\end{flushright}
\end{quote}


\textcolor{red}{Our sensor data from Figure \ref{fig:a_data_description_env_T_WBGT} also shows higher wet bulb globe temperature during monsoon. skin temperature of rickshaw pullers during monsoon is also very high, for some subjects exceeding 35 °C (Figure \ref{fig:a_data_description_env_T_skin})}.

\textcolor{red}{Moreover, thermal stress causes more sweating as part of the thermoregulation. The correlation network in Figure \ref{fig:env_phys} also suggests a significant positive correlation between weather variables and skin conductance level ($SCL$). Complementing this quantitative evidence, many interviewees reported increased sweating and weakness. A 25-year-old participant mentioned, \textit{`I sweat more than before, maybe due to the higher temperature.'}.}

\begin{quote}
\textcolor{red}{\textit{`It doesn't rain on time, sometimes it gets hot unexpectedly, and it gets cold at odd times. Two or three years ago, it wasn’t this hard; now it’s harder. Due to the increase in temperature, I feel weaker, and it affects my health. If it gets too hot, my body lacks energy, which makes me feel tired.'}}
\begin{flushright}
\textit{\textcolor{red}{--- A 25-year-old rickshaw puller, experience driving for 10 years}}
\end{flushright}
\end{quote}

\textcolor{red}{Our sensor data is also showing similar results. Figure \ref{fig:a_data_description_SCL} shows that the skin conductance level (an indicator of sweat-gland activity) stays above 15 $\mu S$ during both riding and recovery.
In Figure \ref{fig:a_data_description_env_SCL} we see that, throughout most of the monsoon and summer seasons, it even exceeded the 20 $\mu S$ threshold that marks extreme sympathetic arousal \cite{7_Boucsein2012}.}


\subsection{\textcolor{red}{A Physically Strenuous Job getting More Difficult}}
\textcolor{red}{According to the interviewees, the combination of climate changes, demographics, and socioeconomic conditions has made the already strenuous job of rickshaw driving even more difficult. All participants identified heatwaves in Dhaka as a recent phenomenon and agreed that extreme heat significantly increases the physical effort required to drive a rickshaw. }

\begin{quote}
\textcolor{red}{\textit{{Interviewer:} Well. Do you think that climate change is making it harder for you to drive your rickshaw than before?\\
{Rickshaw Puller:} Yes. It's becoming harder. It requires more effort in the heat. The heat dehydrates my body.\\
{Interviewer:} You have been driving rickshaw for years. Do you get tired faster now than before?\\
{Rickshaw Puller:} Yes, I definitely tire faster now. I start around 10 to 10:30 AM and get exhausted around 3 to 4 PM.}}
\begin{flushright}
\textcolor{red}{\textit{--- A 59-year-old rickshaw puller, experience driving for 20 years}}
\end{flushright}
\end{quote}

\textcolor{red}{Our collected sensor data, specifically relative cardiac cost ($RCC$) also suggests a similar indication. Figure \ref{fig:a_data_description_env_RCC} suggests that even during winter, $RCC$ exceeded 40\% for some subjects, while Monsoon and Summer data show very close to 40\% and above. Figure \ref{fig:a_data_description_RCC} suggests that during riding, $RCC$ spiked above 40\%, which is a strict categorization of vigorous or heavy effort in occupational literature \cite{5_s150716956}. A similar concern from a rickshaw puller,}

\begin{quote}
\textcolor{red}{\textit{{Interviewer:} Are you getting tired faster nowadays?\\
{Rickshaw Puller:} I start at 7 AM in the morning and continue until 3 to 4 PM. By 11 AM, my body begins feeling tired. If you mention ten years ago, then imagine when there was strong sunlight, after 1-1:30 PM, I used to feel a little tired.\\
{Interviewer:} So your fatigue now arrives two hours earlier.\\
{Rickshaw Puller:} Exactly. Two hours earlier.}}
\end{quote}

\textcolor{red}{Going forward, when discussing the physical challenges of driving rickshaws in cold weather and rain, participants expressed varying opinions. However, nearly all agreed that winter makes the job less physically demanding. Aging is also cited as a factor that increases the physical strain of rickshaw driving. Many participants mentioned that, as they get older, the work feels more taxing. Hence, $Age$ appears as an important predictor in regression models, particularly in modeling $RCC$. }

\begin{quote}
\textcolor{red}{\textit{`It is indeed becoming hard. I started driving rickshaws at the age of twenty. I have been driving rickshaws since before I got married. It wasn’t like this then. Now I am getting older. I go out in the morning, and I start feeling tired after 10 AM. I remain tired until 12 at night. I feel tired when I go home, and I feel tired on the road. There is no rest in between.'}}
    \begin{flushright}
\textcolor{red}{\textit{--- A 52-year-old rickshaw puller, experience driving for 25 years}}
\end{flushright}
\end{quote}


\subsection{Absence of Climate Education and Reliance on Traditional Knowledge and Beliefs}
Many of our participants appear to be unfamiliar with the term climate change and lack formal education on the subject. Coming from lower socioeconomic backgrounds with limited access to formal education, they rarely have the opportunity to attend college. However, when we explain the concept of climate change by providing real-world examples, they demonstrate an awareness of the unusual changes occurring in the environment. Despite their lack of formal education on climate change, all participants acknowledge the obvious gradual shifts in our climate.

\begin{quote}
\textit{`Earlier, during the months of Ashar and Shravan (two months in the Bangla calendar), it used to rain. It was the Rainy (Monsoon) season. Now, the rainy season falls in Bhadra and Ashwin (the months following Ashar and Shravan).'}
\begin{flushright}
\textit{--- A 39-year-old rickshaw puller, experience driving for 13 years}
\end{flushright}
\end{quote}

\begin{quote}
\textit{`Before, we used to have six months of monsoon, followed by winter, and then summer. But now it feels like it’s hot for nine months of the year. The monsoon has shortened, and winter has decreased even more. The temperature during summer is also much higher now.'}
\begin{flushright}
\textit{--- A 58-year-old rickshaw puller, experience driving for 14 years}
\end{flushright}
\end{quote}

Despite their limited access to modern information sources, traditional knowledge has helped bridge some gaps in their scientific understanding of climate change. Some participants reported hearing about climate change or natural disasters through mainstream media like TV, radio, and newspapers. One even mentioned learning about climate change through conversations with passengers.

\begin{table*}[!tbp]
\centering
\caption{Comparison of our study with existing literature.}
\label{tab:discussion}
\resizebox{\linewidth}{!}{
\rotatebox{90}{%
\begin{tabular}{|l|l|l|l|l|l|l|l|l|l|l|l|l|}
\hline
Study & Year & Dataset & Subject & Data type & \begin{tabular}[c]{@{}l@{}}Weather \\ variables\end{tabular} & \begin{tabular}[c]{@{}l@{}}Physiological \\ variables\end{tabular} & \begin{tabular}[c]{@{}l@{}}Activity \\ monito-\\ ring\end{tabular} & \begin{tabular}[c]{@{}l@{}}Wear- \\able \\ sensing\end{tabular} & \begin{tabular}[c]{@{}l@{}}Correla- \\tion \\ analysis\end{tabular} & \begin{tabular}[c]{@{}l@{}}Bayesian \\ Network\end{tabular} & \begin{tabular}[c]{@{}l@{}}Climate \\ model-\\ing\end{tabular} & \begin{tabular}[c]{@{}l@{}}Surviv- \\ability \\ analysis\end{tabular} \\ \hline
\begin{tabular}[c]{@{}l@{}}Vanos \\ et al., \cite{Vanos2023}\end{tabular} & 2023 & \begin{tabular}[c]{@{}l@{}}GFDL ESM4 \\ and \\ MPI ESM1.2\end{tabular} & - & \begin{tabular}[c]{@{}l@{}}Climate \\ and \\ physiology\end{tabular} & \begin{tabular}[c]{@{}l@{}}Air temperature, \\ relative humidity, \\ and \\ wet bulb temperature\end{tabular} & \begin{tabular}[c]{@{}l@{}}Metabolic \\ rate\end{tabular} & - & - & - & - & $\boldsymbol{\checkmark}$ & $\boldsymbol{\checkmark}$ \\ \hline
\begin{tabular}[c]{@{}l@{}}Sher-\\ wood \\ et al., \cite{Sherwood2010}\end{tabular} & 2010 & CAM3 & - & Climate & \begin{tabular}[c]{@{}l@{}}Air temperature \\ and \\ wet bulb temperature\end{tabular} & - & - & - & - & - & $\boldsymbol{\checkmark}$ & $\boldsymbol{\checkmark}$ \\ \hline
\begin{tabular}[c]{@{}l@{}}Luthi \\ et al., \cite{Lüthi2023}\end{tabular} & 2023 & \begin{tabular}[c]{@{}l@{}}CESM1.2, \\ CESM1-CAM5, \\ CanESM2, \\ GFDL-ESM2M,\\ and \\ CSIRO-Mk3.6.0\end{tabular} & - & Climate & air temperature & - & - & - & - & - & $\boldsymbol{\checkmark}$ & $\boldsymbol{\checkmark}$ \\ \hline
\begin{tabular}[c]{@{}l@{}}Ionnou \\ et al., \cite{Leonidas2022}\end{tabular} & 2022 & \begin{tabular}[c]{@{}l@{}}n=2,409\\ \\ Meta\\ analysis\end{tabular} & \begin{tabular}[c]{@{}l@{}}Outdoor \\ workers\end{tabular} & Physiology & - & \begin{tabular}[c]{@{}l@{}}Core and \\ skin temperature, \\ metabolic rate, \\ and \\ urine specific gravity\end{tabular} & - & - & $\boldsymbol{\checkmark}$ & - & $\boldsymbol{\checkmark}$ & $\boldsymbol{\checkmark}$ \\ \hline
\begin{tabular}[c]{@{}l@{}}Taggart \\ et al., \cite{Taggart2023}\end{tabular} & 2023 & n=27 & \begin{tabular}[c]{@{}l@{}}Mine \\ industry \\ workers\end{tabular} & \begin{tabular}[c]{@{}l@{}}Weather \\ and \\ physiology\end{tabular} & \begin{tabular}[c]{@{}l@{}}Air temperature, \\ relative humidity, \\ and WBGT\end{tabular} & \begin{tabular}[c]{@{}l@{}}Core temperature, \\ and \\ urine specific gravity\end{tabular} & $\boldsymbol{\checkmark}$ & $\boldsymbol{\checkmark}$ & - & - & - & - \\ \hline
\begin{tabular}[c]{@{}l@{}}Consta-\\ ntinou \\ et al., \cite{Constantinou2021}\end{tabular} & 2021 & n=41 & \begin{tabular}[c]{@{}l@{}}Population \\ from urban \\ and \\ rural areas\end{tabular} & \begin{tabular}[c]{@{}l@{}}Weather \\ and \\ physiology\end{tabular} & Air temperature & Skin temperature & $\boldsymbol{\checkmark}$ & $\boldsymbol{\checkmark}$ & $\boldsymbol{\checkmark}$ & - & - & - \\ \hline
\begin{tabular}[c]{@{}l@{}}Kamal \\ et al., \cite{Kamal2024}\end{tabular} & 2024 & \begin{tabular}[c]{@{}l@{}}ECMWF \\ ERA5\end{tabular} & - & Weather & WBGT & - & - & - & - & - & - & - \\ \hline
\begin{tabular}[c]{@{}l@{}}Justine \\ et al., \cite{Justine2023}\end{tabular} & 2023 & HadISD & - & \begin{tabular}[c]{@{}l@{}}Climate, \\ weather, \\ and \\ mortality\end{tabular} & \begin{tabular}[c]{@{}l@{}}Air temperature, \\ relative humidity, \\ and \\ wet bulb temperature\end{tabular} & - & - & - & - & - &  & $\boldsymbol{\checkmark}$ \\ \hline
\begin{tabular}[c]{@{}l@{}}Chandan \\ et al., \cite{Chandan_2008}\end{tabular} & 2008 & \begin{tabular}[c]{@{}l@{}}n=106 in \\ laboratory\end{tabular} & \begin{tabular}[c]{@{}l@{}}Rickshaw \\ pullers\end{tabular} & Physiology & - & \begin{tabular}[c]{@{}l@{}}Relative cardiac \\ cost\end{tabular} & $\boldsymbol{\checkmark}$ & $\boldsymbol{\checkmark}$ & - & - & - & - \\ \hline
\begin{tabular}[c]{@{}l@{}}Manna \\ et al., \cite{manna2012physiological}\end{tabular} & 2012 & n=18 & \begin{tabular}[c]{@{}l@{}}Rickshaw \\ pullers\end{tabular} & Physiology & - & \begin{tabular}[c]{@{}l@{}}Energy expenditure, \\ working heart rate\end{tabular} & $\boldsymbol{\checkmark}$ & $\boldsymbol{\checkmark}$ & - & - & - & - \\ \hline
\begin{tabular}[c]{@{}l@{}}Sahu \\ et al., \cite{Sahu2013_tq}\end{tabular} & 2013 & n=142 & \begin{tabular}[c]{@{}l@{}}Rickshaw \\ pullers\end{tabular} & Physiology & WBGT & \begin{tabular}[c]{@{}l@{}}Working hear rate, \\ recovery heart rate,\\ and \\ relative cardiac cost\end{tabular} & $\boldsymbol{\checkmark}$ & $\boldsymbol{\checkmark}$ & - & - & - & - \\ \hline
\begin{tabular}[c]{@{}l@{}}Islam \\ et al., \cite{Islam2016SocioEP}\end{tabular} & 2016 & n=50 & \begin{tabular}[c]{@{}l@{}}Rickshaw \\ pullers\end{tabular} & \begin{tabular}[c]{@{}l@{}}Socio-\\ demographics\end{tabular} & - & - & - & - & - & - & - & - \\ \hline
\begin{tabular}[c]{@{}l@{}}Karim \\ et al., \cite{karim2019organising}\end{tabular} & 2019 & - & \begin{tabular}[c]{@{}l@{}}Rickshaw \\ pullers\end{tabular} & - & - & - & - & - & - & - & - & - \\ \hline
\begin{tabular}[c]{@{}l@{}}Our \\ study\end{tabular} & 2024 & n=50 & \begin{tabular}[c]{@{}l@{}}Rickshaw \\ pullers\end{tabular} & \begin{tabular}[c]{@{}l@{}}Weather \\ and \\ physiology\end{tabular} & \begin{tabular}[c]{@{}l@{}}Air temperature, \\ relative humidity, \\ wet bulb \\ temperature, \\ and WBGT\end{tabular} & \begin{tabular}[c]{@{}l@{}}Relative cardiac cost, \\ skin temperature, \\ and\\ exposure time\end{tabular} & $\boldsymbol{\checkmark}$ & $\boldsymbol{\checkmark}$ & $\boldsymbol{\checkmark}$ & $\boldsymbol{\checkmark}$ & $\boldsymbol{\checkmark}$ & $\boldsymbol{\checkmark}$ \\ \hline
\end{tabular}
}
}
\end{table*}

\section{Discussion}
This study explores three research directions: understanding the changes in physiological biomarkers experienced by rickshaw pullers due to heat exposure; leveraging weather variables, activity features, and climate models to forecast physiological biomarkers in the future; and analyzing survivability in the future based on widely used metrics and thresholds. 

\subsection{RQ1}
To answer the first research question, we construct a dataset based on 100 rickshaw pullers. In contrast to previous studies, our dataset is unique in several respects. First, it is collected under diverse weather conditions spanning three different seasons. Second, it integrates multiple dimensions of occupational heat exposure, aggregating weather variables, demographic information, activity levels, and physiological biomarkers. In Table \ref{tab:discussion}, we summarize related studies in this domain and highlight our contributions across these criteria, while also acknowledging existing limitations.

Among the limited studies on rickshaw pullers, only Sahu et al. \cite{Sahu2013_tq} developed a dataset of 142 participants, focusing on wet bulb globe temperature (WBGT) and relative cardiac cost. Other studies either examined only physiological variables, without incorporating weather conditions \cite{Chandan_2008, manna2012physiological}, or analyzed solely the socio-demographic characteristics of rickshaw pullers \cite{Islam2016SocioEP, karim2019organising}. Compared to our study, these works are constrained by the absence of integrated weather variables, physiological biomarkers, and modeling of future exposures.

Based on our dataset, we then construct correlation networks. The networks and the corresponding correlation matrix in Table \ref{tab:correlation_matrix} address the first research question, which seeks to understand how physiological biomarkers are associated with other variables. The findings indicate that real-time weather variables ($T_{air}$, $R_H$, and $T_{WBGT}$) are the most influential in assessing the physiological impacts on rickshaw pullers.

In comparison to our approach, some existing studies performed statistical hypothesis testing on rickshaw pullers' physiological biomarkers to understand the influence of demographic factors (e.g., age, weight, and location) \cite{Chandan_2008, manna2012physiological, Sahu2013_tq}. 
However, most of the existing studies worked with limited-scale physiological variables and weather data. 

\subsection{RQ2}
Our findings from the correlation analysis encourage us to explore activity, demographics, and weather in modeling physiological biomarkers. The Linear Gaussian Bayesian Network provides more interpretation in understanding the cause-and-effect relations in this domain. The structure of the network is learnt mostly from the data itself, making it more reliable to model the physiological biomarkers. Instead of other regression models, this network can be better used to analyze the effect of a single variable or a set of variables on the network. 

\textcolor{red}{Overall, all physiological variables except RCC achieve correlation coefficients above 0.65 between the predicted and ground-truth values. For RCC, an MAE of 9.02 (range 6.4–63) yields a moderate NMAE of 0.169. The primary limitation in modeling RCC with better performance metrics stems from multiple reasons. First, cardiac cost is highly influenced by other factors, such as the mechanical efficiency of rickshaws, the weights of passengers, etc. These factors pose a higher variability in cardiac output. Besides, A related study on physiological load in logging workers examined several predictors of cardiovascular strain (RCC) \cite{ijerph19137695}. They also reported correlation coefficients between RCC and features ranging from 0.14 to 0.15, which is pretty similar to our feature correlations for RCC (ranging from 0.15-0.26) in Table 3. Because of such moderate correlations, even present in the literature, it is challenging to model RCC with better performance.

\subsection{RQ3}
The geographic region of Bangladesh shows an increasing trend in surface temperature and relatively stable trends in relative humidity and solar radiation, as indicated by global climate model ensembles (Figure \ref{fig:climate_model_inference_environment}). In this context, assessing the survivability of rickshaw pullers under both current (2023–2025) and projected climate conditions is critical. Using two key metrics: $T_{WBGT}$ and $T_{skin}$, as defined in Table \ref{tab:survivavility_thresholds}, we evaluate how survivability may change in future decades. As shown in Table \ref{tab:compare_surviabailiy_present_future}, the results suggest that an increasing proportion of rickshaw pullers will be exposed to extreme heat conditions in the coming decades. These findings are consistent with Ioannou et al. \cite{Leonidas2022}, who project that physical work capacity could decline to below 50\% by the summer of 2030 in certain regions of the world, including Bangladesh and other South Asian countries such as India, Sri Lanka, and Pakistan.

Further, Thematic analysis of interview data reveals several key themes, some of which corroborate findings from the sensor-based quantitative analysis.
First, participants widely acknowledged shifts in weather patterns, even among those with limited awareness of climate change. This perception aligns with our sensor data from 2023–2024. Many rickshaw pullers reported observing these changes over the past 5–10 years.
\textcolor{red}{Second, the two dominant themes: (i) Heat Intensity and Thermal Stress and (ii) A Physically Strenuous Job getting More Difficult, highlight the vulnerability of rickshaw pullers even in the current climate conditions. Many find the occupation getting strenuous compared to earlier years. These observations align with recent news articles on rickshaw pullers' demand towards moving to electric-powered rickshaws \cite{dailystar_rickshaw, tbs_rickshaw}.}


\section{Limitations and Future Directions}

Our study has several limitations. 
A primary limitation is that our dataset consists exclusively of male rickshaw pullers, reflecting the sociodemographic context and the physically demanding nature of this occupation. This demographic and regional concentration limits the generalizability of our findings to female workers, individuals in other occupations, and populations in different geographic settings. 

Another limitation is the assumption that rickshaw pullers will maintain their current work schedules (e.g., trip timing, routes, and speed) under future climate conditions. However, they may adapt by altering their schedules or transitioning to other professions if climate conditions become intolerable.

\textcolor{red}{ A further limitation is that, except for $RCC$, all physiological variables demonstrate better performance in regression. Considering the challenges in modeling RCC \cite{ijerph19137695}, we emphasize a continuous effort to collect more data in the future. }}

\textcolor{red}{An additional concern relates to the temporal scale mismatch between wearable data (minutes) and climate projections (monthly). Primarily, temporal scale mismatch can introduce smoothing of extreme events. Climate projections can smooth out short-term extreme events, such as heat waves, that can significantly impact physiological responses. On the other side, CMIP6-based models assume a uniform monthly offset, treating all times of day equally. However, asymmetric future warming (e.g., greater increases in temperature during daytime than nighttime) can underestimate peak daytime heat stress, biasing projections of physiological variables. }

\textcolor{red}{Finally, CMIP6-based global climate models do not capture fine-grained urban microclimates relevant to rickshaw routes. Though we try to simulate a plausible future climate scenario by integrating climate forecasts with sensor data, this method is not fully physics-informed and does not capture dynamic urban factors such as the shade of buildings and trees, narrow road segments, reflection of thermal radiation from street surfaces, etc. Such global climate models smooth over the mentioned microclimate `hotspots' where rickshaw drivers face the highest risks. Therefore, there is a necessity for future work to integrate high-resolution urban microclimate models to improve spatial fidelity.}


\section{Conclusion}
In this study, we prepare a heat exposure dataset (n=100) for rickshaw pullers. Then, we build a climate forecast-based regression model. The model can predict how physiological biomarkers (e.g., skin temperature, skin conductance response, skin conductance level, and relative cardiac cost) may look in the future under different climate scenarios. Using the model, we analyze the biomarkers in present and future climate scenarios. To validate our findings, we conducted interviews with 12 rickshaw pullers, asking about their perception and experience of climate change in their workplace. Through this mixed-method study, we offer an exploration of the rickshaw pulling experience, both in the present and under potential future climate scenarios.

\begin{acks}
We sincerely thank the rickshaw pullers in Dhaka, Bangladesh, for generously sharing their time and experiences, without whom this work would not have been possible. We are grateful to the Merkin Graduate Fellows Program and the Halıcıoğlu Data Science Institute (HDSI) at UC San Diego for providing wearable and environmental sensors, as well as participant compensation. We sincerely acknowledge the support of the Bangladesh University of Engineering and Technology (BUET). Besides, we extend our deep appreciation to the BUET undergraduate students whose dedicated efforts in data collection were instrumental in preparing the dataset and making this research possible.
\end{acks}


\bibliographystyle{ACM-Reference-Format}
\bibliography{sample-base}

\appendix

\end{document}